# Principles of the Concept-Oriented Data Model

Alexandr Savinov

Institute of Mathematics and Computer Science
Academy of Sciences of Moldova
Academiei 5, MD-2028 Chisinau, Moldova

Fraunhofer Institute for Autonomous Intelligent System
Schloss Birlinghoven, 53754 Sankt Augustin, Germany

http://www.conceptoriented.com
savinov@conceptoriented.com

# Principles of the Concept-Oriented Data Model

Alexandr Savinov

Institute of Mathematics and Computer Science, Academy of Sciences of Moldova
Academiei 5, MD-2028 Chisinau, Moldova

Fraunhofer Institute for Autonomous Intelligent System
Schloss Birlinghoven, 53754 Sankt Augustin, Germany

http://www.conceptoriented.com
savinov@conceptoriented.com

In the paper a new approach to data representation and manipulation is described, which is called the concept-oriented data model (CODM). It is supposed that items represent data units, which are stored in concepts. A concept is a combination of superconcepts, which determine the concept's dimensionality or properties. An item is a combination of superitems taken by one from all the superconcepts. An item stores a combination of references to its superitems. The references implement inclusion relation or attribute-value relation among items. A concept-oriented database is defined by its concept structure called syntax or schema and its item structure called semantics. The model defines formal transformations of syntax and semantics including the canonical semantics where all concepts are merged and the data semantics is represented by one set of items. The concept-oriented data model treats relations as subconcepts where items are instances of the relations. Multi-valued attributes are defined via subconcepts as a view on the database semantics rather than as a built-in mechanism. The model includes concept-oriented query language, which is based on collection manipulations. It also has such mechanisms as aggregation and inference based on semantics propagation through the database schema.

## 1  Introduction

Currently there exists a wide spectrum of models used to represent data, which are based on different basic principles. These principles lie in different planes so that there exist different dimensions used to distinguish data models. For example, the models could be distinguished by the formalism used to describe the data, by query language used to retrieve data, by data manipulation operators provided by the model, by the model storage principles etc. Most existing models are not isolated and intersect with other models on one set of principles while sharing another set of principles. Thus there exist different classification schemas that can be used to distinguish them.

Although there exist different alternative sets of criteria for classifying data models, we can always select the most important principles having the highest level of abstraction. Frequently these principles are so general that cover other disciples rather only data models and databases. In many cases however, these principles are not declared explicitly so that it is quite difficult to reconstruct them and unambiguously classify one or another data model. This abstract set of principles determines the whole paradigm a data model is based on, i.e., they establish a high level view of the world. Normally these principles concern such issues as what is data, what is data representation, what is relation, what is access, what is operation, what is syntax and semantics, what consistency and constraints, what is inference and so on.

The data model described in this paper is based on the concept-oriented paradigm, which is based on its own fundamental principles. The paradigm is much wider than the data model and underlies other areas such as programming, modelling or analysis and design. In this sense it is like object-oriented paradigm that is based on one set of general principles used in different disciplines such as programming or modelling. Some concept-oriented principles are directly used in the described data model while other are have weak consequences for this area. Some principles are rather concrete and constructive while other have in great extent philosophical character. Many of these principles are not independent and may have different level and priority. Yet below in the introduction we try to formulate all of them explicitly as independent highest priority statements or axioms because without them it is quite difficult to understand the essence of this approach and its difference from other paradigms in general and data representation methods in particular. Indeed at the molecular or atomic level of databases are simply elementary interactions so they are all equal. At the level machine language they still equivalent. And even at the level procedural programming most of the data models cannot be distinguished. It is only a set of high level organisational principles that can help us to distinguish one model from another.

The first principle specifies the main areas of interest of the concept-oriented paradigm:

> **CO1** The concept-oriented paradigm is aimed at studying representation and access issues in any system.

Thus representation and access are the key words for this direction (interaction could be added to this list). In other words, for any system we are first of all interested in how things are represented, how they are access and how they interact.

The next principle introduces types of things used in our analysis:

> **CO2** There exists two sorts of things: objects and spaces.

These things have numerous synonyms used in different contexts and formal settings. For example, object are also called entity, item, value, element, record etc. Space is can be called set, class, concept, domain etc. In the concept-oriented data model space is associated with the model syntax (section 2.1) while objects are associated with the model semantics (section 2.2). The separation between these two sorts of things is not unique for the concept-oriented paradigm and essentially underlies all the contemporary mathematics (after Descartes) where the world is described by means variables and values. However, in the concept-oriented paradigm it is explicitly formulated as one of the highest priority principles. In this principle we simply recognize that there exists a deep fundamental difference between these two sorts of things without which we are not able to describe the world in general and represent data in particular. Of course much better would be to get rid of this separation and reduce all operations to only one type of things. However, it is currently impossible because the whole world outlook is based on this assumption. In other words, in order to change that we need a new vision of the world organisation, which is a kind of the theory of everything where any entity can be described by means of itself. We do not exclude such a theory at all but for the needs of the data representation we assume that this principle is true. By postulating the existence of two sorts of things we essentially make our life easier. In particular, we can formulate the next principle:

> **CO3** Objects are living in space and its structure determines most of the whole system functionality, which is concentrated on the space borders.

Actually here we combined three statements: (i) objects are living in space, (ii) the space structure determines the system functionality, and (iii) the space functionality is concentrated on its borders. The first statement establishes the major relationship between objects and spaces; namely, we say that the space is a primary thing that exists by itself while objects are not able to exist without the corresponding space. Why objects need space for their existence? Because space provides an environment with necessary functionality. If the space has a complex structure then the functionality is also complex. Here we come to the second statement, which postulates that the complexity of large systems is due to their complex space structure. This is already unique for the concept-oriented paradigm. Compare it with what the object-oriented paradigm says: the system functionality and complexity is determined by the objects. Indeed, following such a object-oriented approach in order to build a complex system we have to describe complex objects. In the concept-oriented approach we essentially do not care if objects are complex or simple because we say that it is the space that is responsible for the system functionality. In other words, if you want to develop a system then according to (ii) first develop its space structure with the corresponding functions and only after that put objects into this space (statement (i)). The third statement simply says that this functionality is concentrated on space borders so it is activated when processes intersect these borders. In general case the system functionality is distributed between objects and space structure and both types are considered in the concept-oriented paradigm. However, the development of space structure and its functions is more important especially for complex systems and it is an original part of the concept-oriented paradigm. This principle is more important for the concept-oriented programming while for data modelling (in the form described in this paper) it is almost not used.

The next issue we would like to clarify is what is the space and what are objects in it. The concept-oriented treatment of the space structure and object existence is declared in the following principle:

> **CO4** The space has a multidimensional and hierarchical (multilevel) structure and object existence is spread over the levels and dimensions.

This principle also involves several statements. The first one restricts the space structure by a set of multidimensional hierarchical spaces. This means that any system has different dimensionalities and can be viewed from different (orthogonal) sides. Simultaneously the system has levels and can be viewed or described at different levels of details or abstraction. Both these properties of the space structure are of extreme importance for the concept-oriented paradigm. In particular, a complex system cannot be reduced to the collection of elements or to the combination of elements. The only way consists in representing it by both methods. Such a structure imposes a strong constraint on the system organisation but such a self-restriction is supposed to be compensated by an ability to clearly describe any complex system.

The second part of this principle claims that object identity or existence is spread over the space structure. The importance of this rule becomes clear if we ask the question what an object is. For example, is a record equivalent to its primary key, is equivalent to its physical position in the storage? This rule says that both representations are equivalent, i.e., any object may accept any form depending on the level and dimensions used to represent it. This is a kind of general law, which can be used in much wider scope. For example, is a person his passport, his social security card, his photo or recollections of relatives? The rule says that all of them can be considered the person representation. Moreover, until these evidences exist at least in some form on some level the object is considered existing. One consequence is that we never know what object is reality because the reality does not exist, i.e., it does not have the ultimate representation. So that question what is the record's ultimate representation does not make sense much sense because on different levels it has different representation. For example, it could be considered a set of physical interactions in semiconductor or a set of bits in memory. Yet the situation is not so uncertain. The thing is that objects still have a unique identity and we can always connect different representations because the space has a concrete structure and all representations follow it.

The next principle defines more precisely what is meant by multidimensional hierarchical structure of space:

> **CO5** Space is a combination of other spaces.

Objects are defined similarly in the following principle:

> **CO6** Object is a combination of other objects.

In different areas where the concept-oriented paradigm can be applied these definition can be made more specific. In particular, in the concept-oriented data model we prohibit loop (both for spaces and objects) and require for objects to combine other objects taken from its direct subspaces. However, the most interesting and extremely important for data modelling is how the principle CO6 can be interpreted. It turns out that this simple principle can solve many fundamental modelling problems especially in the area of data modelling. In particular, the relationship between objects arising from this principle can be interpreted as:

**CO6.1** Coordinate assignment and positioning objects in space. Here we suppose that each object constituents (i.e., a set of other objects that are combined in this object) are simultaneously its coordinates in the corresponding spaces. Notice that the coordinates themselves are also objects with their own coordinates in other spaces.

**CO6.2** Attribute value assignment and object characterization. Here we suppose that object constituents are its characteristics or attribute values (spaces are attribute domains). Notice that the values are themselves normal objects with their own characterization.

**CO6.3** Object inclusion into other objects. Here we suppose that each object is included into each of the objects it combines. The object constituents are then the sets collecting many other objects because they are involved in many combinations.

**CO6.4** Mechanism of multi-valued attributes. Each item can be included into many combinations simultaneously, which can be then considered its multi-valued attributes.

These are only the most important consequences of the definition of object as a combination of other objects. Interestingly such a definition assumes that objects cannot be defined without other objects and exist in isolation from them. Objects in the sense of this definition simply do not have their own distinguishing features or characteristics and the only way to provide some semantics for them is to connect them with other objects. In the set of secondary principles CO6.* we simply provided the most important in the context of this paper interpretations of such connections. Notice that here we managed to avoid "multiplying entities beyond necessity" because we do not separate types of entities for positioning and coordinates, attributes and values, inclusion of objects into sets, multi-valued attributes and many other more specific mechanisms. All these things are defined via the main principle CO6.

We generally outlined what is meant by objects and spaces in the concept-oriented paradigm However, there is still one crucial element of any system, which needs to be defined — it is a relation. Indeed, putting objects into their positions in space is not enough. Objects interact not only with the structural elements of the space but also with each other. In other words, objects have some information about other objects and this information can be interpreted as relations between them. The main difficulty here is that relations are not objective, i.e., they cannot be treated as objects because they are somewhat virtual elements. It is something imaginable because we normally cannot identify them and qualify as real objects. For example, what is the neighbourhood relation between objects in physical space? If it is not an object then what is it? If it is an object then where does it live? In order to separate spaces and objects we introduced two sorts of things (for the purposes of this paper). This (bad) solution could be applied also for the relations. However, in this case we would not like to make a compromise and are not going "to multiply entities beyond necessity". Instead we formulate the following principle:

> **CO7**   Relations (instances) are objects of lower levels.

This principle essentially says that relations do not exist and their instances are normal objects. Thus objects and relation instances are not distinguishable taken by themselves in isolation from other elements of the system. This role becomes relative, i.e., an object is treated as a relation for the objects of higher levels while for objects of lower levels (which are relation instances) it is a normal object. This explains why relations are not visible and cannot be viewed as normal (corpuscular) objects. This is because they exist on lower levels. Here it is important to note the existence of hierarchical structure in any system postulated in the principle CO4. Without such an internal hierarchical space the system would not be able to have relations.

We described the most important principles, which are common for the whole concept-oriented paradigm. We used very the conventional terms such as space and object in order to emphasize rather general character of this theory. However, since the concept-oriented approach is a separate area with its own goals and distinguishing features later in the paper we will use special terms for the main notions. In particular, we will use the term concept instead of space and the term item instead of object. Yet when we need to emphasize the general nature of some statement we will still use the terms space and object. Also we will use other equivalent terms like record or set in order to emphasize an analogy or similarity with other areas such as the set theory or the relational model.

In section 2 of the paper we formally introduce the main notions and terms of the concept-oriented data model. In section 3 the concept-oriented query language is described. In section 4 different more specific mechanisms and modelling techniques are described. So essentially this section is intended to demonstrate the practical value of this approach and its ability to model different cases. In section 5 we describe the relationship to other main data modelling methods.

# 2 Data Representation

## 2.1 Database Syntax

The main notion in the concept-oriented data model is that of *concept*. It substitutes the notion of relation or table in relational databases. Concept is also an analogue of class in programming.

A concept $C$ is characterized by its *intent* $\mathrm{Dim}(C)$ and *extent* $\mathrm{Ext}(C)$.

$$C = \langle D, E \rangle, \text{ where } D = \mathrm{Int}(C) \text{ and } E = \mathrm{Ext}(C)$$

The intent is a combination of *variables* associated with the concept called also attributes, properties, features or slots (Figure 1):

$$D = \text{Int}(C) = \langle x_1, x_2, \ldots, x_n \rangle$$

The number of variables in the intent is said to be the concept dimensionality.

The extent is a set of *items* associated with the concept called also instances of the concept or objects.

$$E = \text{Ext}(C) = \{o_1, o_2, \ldots, o_m\}$$

Frequently concept will be associated with its extent, i.e., we will suppose that concept is a collection of its items: $\text{Ext}(C) = C$. For example, we might write $c \in C$ instead of $c \in \text{Ext}(C)$. Intent and extent are dual notions and this is why we write variables as a combination $\langle \cdots \rangle$ and items as a collection $\{\cdots\}$ of elements. Yet for the sake of simplicity we will frequently write intent as a set or collection $\{x_1, x_2, \ldots, x_n\}$, e.g., by writing $x \in \text{Int}(C)$.

In terms of relational databases concepts are tables or relations, their dimensions are columns and instances represent rows or records. In object-oriented terms concepts are classes and their instances are objects.

There exist a set of predefined *primitive* concepts, which can be used to define new concepts. Primitive concepts are supposed to have no dimensions.

Each dimension in a concept has the corresponding *domain* or range, which is another concept:

$$\text{Dom}(d) = D, \text{ where } d \in \text{Dim}(C) \text{ and } C \text{ and } D \text{ are concepts}$$

The domain determines a set of values the concept instances may take in this dimension.

A *link* is a pair of two concepts where the second concept specifies a domain for the first one:

$$L = \langle C, D \rangle, \text{ where } \text{Dom}(d) = D \text{ and } d \in \text{Dim}(C)$$

Each link can be associated with the corresponding dimension of the source concept.

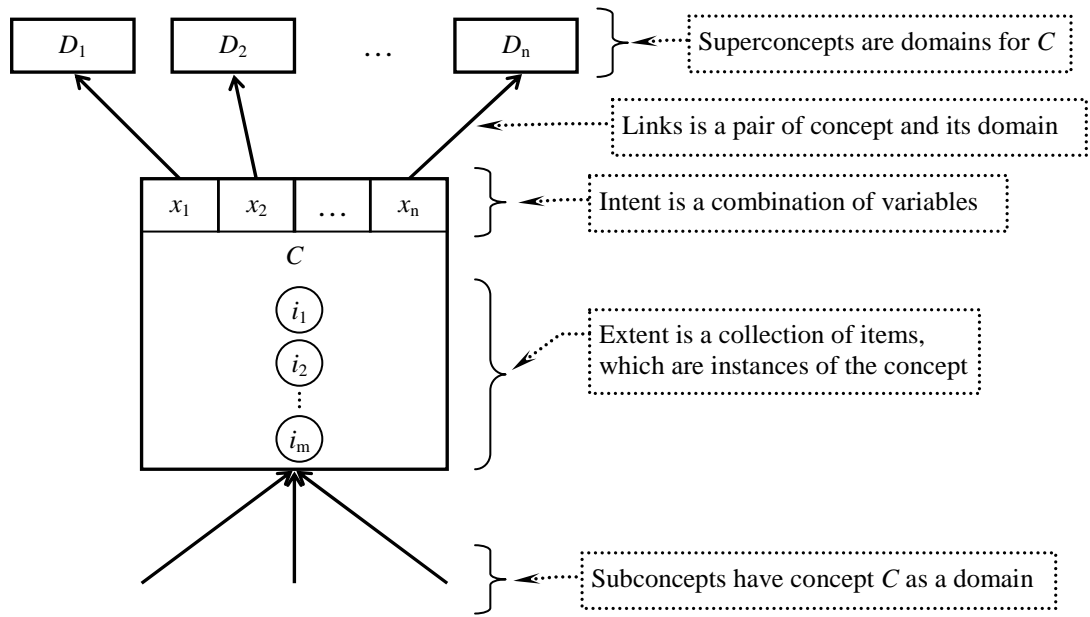

*Figure 1. A concept intent is defined via a combination of variables each with its own domain in some other concept.*

A concept-oriented database *syntax* or schema is a set of concepts with links composing a lattice (Figure 2). Thus database syntax or schema is defined by all intents, i.e., all variables with links to their domains.

A database syntax or schema can be represented by a graph where nodes are concepts and edges are links. Notice that cycles are not allowed in concept-oriented database schema. Strictly speaking loops are also not allowed, i.e., a concept may not have a domain in itself. Yet in practice it can be allowed.

A direct *superconcept* or parent concept is a domain for some dimension of this concept. A direct *subconcept* or child concept is a concept that has this concept as a domain for some its dimension. Thus each concept has a set of direct parents defined by dimension domains and a set of child concepts defined by all concepts with their domain in this concept.

It is convenient to introduce two auxiliary concepts. A *top* or empty concept is a parent for all other concepts, with no its own dimensions. The top concept is associated with empty set and is essentially equivalent to a single item, which could be called Thing, Object or Entity. This item is a superitem for all other items in the database. If a concept does not have an explicitly specified parent then the top concept is used for that purpose. We suppose that all primitive concepts are direct children of the top concept, i.e., the empty concept is a parent for all primitive concepts. Also the top concept can be associated with null value the role of which is described later in the paper.

A *bottom* or full concept is a child for all other concepts in the schema. If a concept does not have a child concept explicitly specified in the schema then the bottom concept is used as its child. The bottom concept is a child for all concepts, which are not used as domains for any dimension in the database schema. The bottom concept is a set, which directly or indirectly includes all items in the database so it is a full set.

A *path* in the schema consists of a sequence of variables starting from a source concept and ending with a target superconcept: $P.x_1.x_2.\cdots.x_k$, where $P$ is the source concept and each variable in the sequence belongs to the intent of the concept pointed to by the previous path: $x_i \in \mathrm{Dim}(P.x_1.x_2.\cdots.x_{i-1})$, $k$ is the length or rank of the path. Path of length $k$ points to (ends with) a superconcept of rank $k$. In particular, path of length 0 is the source concept (with no variables at all) while path of length 1 points to a direct superconcept. Thus each path connects a subconcept with some its superconcept following a sequence of links. Notice that there may be several different paths between two concepts including a child and its direct parent.

A concept own syntax is determined by the variables its intent consists of.

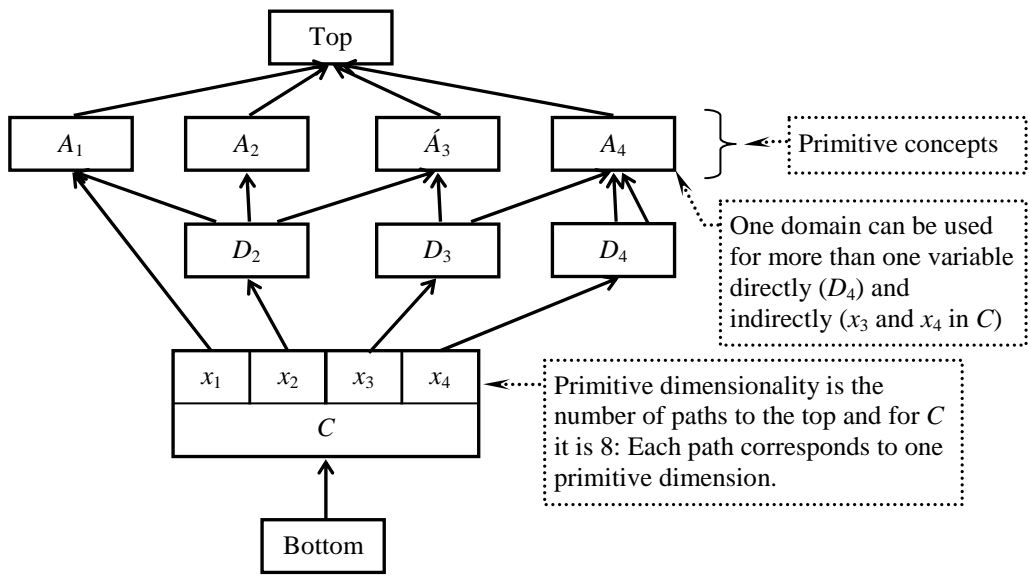

*Figure 2. A database schema is a directed acyclic graph where nodes are concepts and edges are dimensions.*

A concept *primitive* syntax (called also full or canonical) is determined by all paths from this concept to one of the primitive concepts. (In fact, since there is a single top concept this number is equal to the number of paths to the top concept, i.e., the syntax can be defined as all different paths between this and the top concept.) Thus canonical syntax is a combination of all primitive variables that can be reached from this concept following some path: The primitive syntax can be obtained from the concept syntax by recursively substituting domain syntax instead of individual variables till the primitive domains are reached. The canonical syntax consists of only variables with primitive domains.

The database primitive dimensionality (called also full or canonical dimensionality) is that defined by the bottom concept. In other words, the database full dimensionality is the number of paths from the bottom concept to the top concept where each path starts from one of the bottom concept variables and then continues by variables from the new subconcept till one of the primitive concepts (or the top concept) is reached.

## 2.2 Database Semantics

An item or instance of a concept is a combination of other items taken from all the domains of this item concept (Figure 3):

$$i = \langle o_1, o_2, \ldots, o_n \rangle \in C, \ o_i \in \mathrm{Dom}(d_i), \ d_i \in \mathrm{Dim}(C)$$

Thus in order to create a new item for a concept we have to select a single item from each domain associated with this concept.

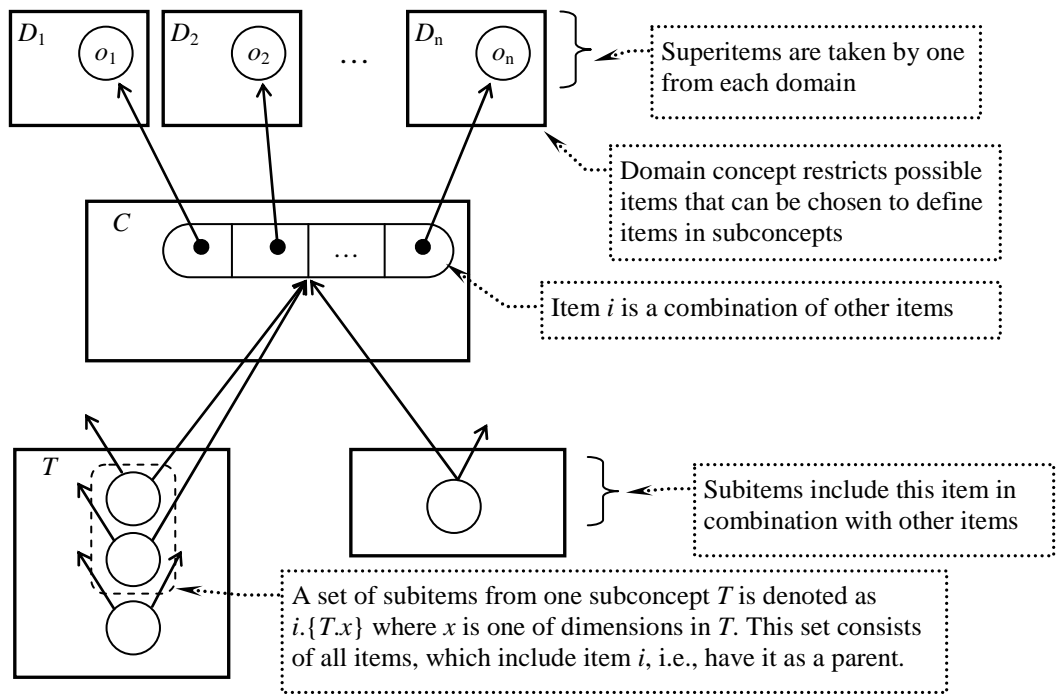

*Figure 3. An item is a combination of other items taken by one from each domain.*

Each instance in concept is represented by its *reference*. In order to store or pass an item we have to use its reference. In most cases item references are generated and managed automatically. However, in some cases custom references can be defined to represent concept instances. It is supposed that reference values are unique with respect to their respective concept.

Some concepts may represent their items by value. In this case the whole item is stored by copying its contents. Such an approach is normally used to represent items from primitive concepts. Thus instances of primitive concepts do not need to be defined and represent themselves by value so that their references are equal to their value.

For any concept we can specify if its items are represented by reference or by value. In most cases however we will suppose that concept instances are represented by references since representation by value or by custom reference is a special mechanism used only in rare situations.

A direct *superitem* or parent item is an instance of the superconcept referenced from this item as a value of one of its dimensions:

$i.d \in T$ is superitem for item $i \in C$, where $d \in \text{Dim}(C)$, and $T = \text{Dom}(d)$

An item is characterized by a set of superitems each taken from exactly one superconcept (or null if allowed). The number of superitems (including nulls) is equal to the number of dimensions. It is important that only a single superitem from one domain can characterize this item so all variables are one-valued and multi-values are not supported. How multi-valued variables can be implemented is described later in the paper in section 4.3.

A direct *subitem* or child item is an instance of some subconcept, which references this item. An item is characterized by several sets of subitems each consisting of subitems taken from exactly one subconcept:

$i.\{C.d\} \subseteq C$ is a set of subitems for item $i \in T$, where $d \in \text{Dim}(C)$, and $T = \text{Dom}(d)$

A direct superitem and subitems are supposed to have level or rank 1.

A superitem of arbitrary rank is defined as a nested relation, i.e., a superitem of rank 2 is a superitem of superitem. A superitem of rank $r$ is an instance of superconcept of rank $r$ referred from this item by means a sequence of $r$ dimensions the first belonging to this concept and the last dimension having the last concept as a domain.

Superitems will be denoted as a sequence of dimensions applied to this item. For example, $i.u.v$ is a superitem of rank 2 for item $i$.

A subitem of arbitrary rank is also defined as a nested relation. A subitem of rank $r$ is an instance of subconcept of rank $r$, which references this item by means of a sequence of $r$ dimensions the first belonging to the subconcept of rank $r$ and the last dimension having this concept as a domain.

Subitems will be denoted by their subsets taken from one subconcept. We will write first this item and then the subconcept and sequence of dimensions leading to this concept. For example, $i.\{C.u.v\}$ is a set of subitems of rank 2 from subconcept $C$, i.e., these are items from $C$, which are characterized by item $i \in T$ via dimensions $u$ and $v$: $i.\{C.u.v\} = \{c \in C \mid C.u.v = i \in T\}$.

Concept own semantics is determined by a set of items its extent consists of.

Concept *primitive* semantics (called also full or canonical) is a set of items consisting of only primitive values corresponding to all primitive dimensions of this concept. Thus each item in canonical semantics is a combination of primitive items that are reached from the source item value following references. The primitive semantics can be obtained from the concept semantics by recursively substituting domain items by value instead of their references till the primitive values are reached. The canonical semantics includes only items consisting of primitive values.

The database primitive semantics (called also full or canonical semantics) is that defined by the bottom concept. It is supposed for the bottom concept that its items are inherited from the direct parent concepts, i.e., for each item from each parent concept there is one item in the bottom concept which references its parent item and has all other variables nulls. Particularly, the number of items in the bottom concept is equal to the sum of items in all its parent concepts.

In fact such a formal inheritance of items does make sense for all concepts but in practice it is not convenient and we assume that all items appear only if explicitly specified in the concept and it is the bottom concept that inherits its items and has no its own items. In other words, in formal analysis it is convenient to suppose that a concept includes all elements of its superconcepts automatically. Such items do not intersect on their dimensions, i.e., this is a simple sum of subsets of items where each subset is characterized by its own dimensions. The concept own items describe what cannot be represented by such a sum, i.e., it includes items that do not fit into individual domains because have more dimensions in their combination of values. This is precisely what makes the semantics non-trivial and further we consider only such non-inherited items as a part of any concept extent. Only for the formally introduced bottom concept it is supposed that its extent is inherited from its domain concepts.

The database canonical semantics (called also complete or primitive semantics) is that defined by the bottom concept.

## 2.3 Item Characterization

In the concept-oriented database model semantics is represented by characterizing new items by a combination of already existing items, which in turn must be characterized by other existing items. The primary role in this process is played by the *characterization* relation between items. This relation can be interpreted as an ability of items to have properties in the form of other items. Notice that property values are normal items just like the characterized item itself. Thus if we have a set of items then its semantics is expressed in the form of this type of relationship between them. If an item has one set of relationships with other items then it has one meaning while changing these relations will result in different interpretation of this item. Thus strictly speaking an item does not have its own semantics or meaning because its semantics is defined via its relationships with other items. In this sense an item cannot be added by itself because adding an item means connecting it with other items. An item cannot exist without other items and connections with them represented in the form of characterization relation. The natural requirement in such an approach is that items cannot characterize one another so the cycles are not allowed and the mutual characterization process is reduced to a set of pre-existing primitive items with no characteristics. For example, we say that a product item is characterized by a producer item, which in turn is characterized by country of origin item.

The process of characterization always starts from an initial set of primitive pre-existing items and then new items are added by specifying their characteristics by choosing a subset of existing items. Such a procedure is interesting from theoretical point of view but is not very convenient for practical tasks. One problem is that each new item can be characterized by almost any already existing item even if we know that it does not make sense. In other words, the potential semantics for each new item is too wide and there is no means that can be used to constrain it. In order to overcome this difficulty we group similar items into groups called concepts. The similarity of items is based on their ability to be characterized by only items from the specified groups of items rather than by all existing items. In other words, items in one concept are known in advance to be characterized by only items from a set of other concepts. On the other hand we know for the items from one concept that they are not characterized by other items than those specified in the concept definition.

Thus in order to describe a new concept we have to specify a set of its characteristics as a set of other already existing concepts. These characteristics or dimensions restrict a set of items that can exist in this concept by only those having their properties as items from the specified dimensions. Notice that the concepts do not define semantics of the data, i.e., they do not represent data itself. Instead they provide dimensionality structure or syntax structure, which can be used to further define semantics by specifying items and their characteristics. After this structure is defined the items are defined always within some concept rather than in global context where they exist with all other items. In other words, for any new item we have to specify its concept and then a number of items from superconcepts that characterize it as properties. Once a concept for an item has been indicated we have a natural constraint for choosing possible properties.

The procedure of characterizing items by other items chosen by one from some set of concepts is analogous to assigning values to object attributes. Indeed, each characterizing concept can be viewed as a domain for the corresponding attribute and choosing a characterizing item from this domain is essentially an assignment of some value. Thus all items of one concept are characterized by a set of attributes and this means that they may take value from the corresponding domains.

An important thing is that an item can be simultaneously an object characterized by some values of its attributes and a value that is used to characterize other items. In other words, we cannot distinguish between characterized objects and characterizing values. Such a relative role of each item and each concept is an important distinguishing feature of the described method of data representation and one of its major design goals. For example, when we describe a concept of car brands by specifying its possible attributes it is a new set of objects. But when we use this concept of car brands to characterize other objects it is already a set of values. In this sense the general ides is that data representation consists in creating items to be used further to describe future new items.

The purpose of concepts consists in providing a complex space or container structure where items are supposed to live. Each concept restricts its items to only a subset of attributes where one attribute is defined as another already existing concept. If an attribute is not specified for a concept then its items are not characterized by its values. The structure of concepts allows us to develop hierarchical dimensions for describing items. Thus items are living in a hierarchical multidimensional space and are characterized by values, which themselves are complex items.

It should be emphasized that an item in concept-oriented database model cannot be defined and exist by itself. The only way to define an item is to specify a combination of other items it consists of. An item does not have its own properties or characteristics, which distinguish it from other items. The semantics of any item is defined exclusively via relationships with other items. In this sense items cannot be defined and exist in isolation from other items. In particular, any property values are also items, which have their own definition via connections with other items. In other words, an item position is space is specified via other items, which have their own position also specified by means of other items and so on until primitive elements are reached. Such a hierarchical multidimensional coordinate system can grow and new elements can be used as new coordinates or positions in space. If we change the meaning (relative position) of an item then this indirectly affects the position of all items that have this coordinate. Notice again that coordinates do not differ from objects they characterize because they themselves are objects with some coordinates.

It should be noticed that the selection of a certain common space structure in the form of concept graph is a necessary element of any knowledge and data sharing. In other words, the common concept structure is used as a static part of description, which is common for many different users or agents and allows for sharing the semantics. Such a knowledge sharing is especially important for ontology development and semantic web where having common terms is one of the necessary elements for creating successful applications. In this case items can be transferred between agents (including databases) and their meaning is not changed and can be easily reconstructed without the necessity to send a set of terms themselves. It is possible because the agents share the top of the database, which contains the common terms while having different bottom parts with their specific semantics. For example, we might standardize the structure of manufactured items and this will guarantee that we can further refer to those items categories in new specific item descriptions and any agent will understand what this item means. In this sense higher level general concepts, which are closer to the top concept should be simultaneously more stable because they define the structure or terms used to describe lower level more specific items. This information plays a role of the basic coordinate system, which should not change very frequently. Lower level concepts, which are closer to the bottom concept may change more frequently because they are final semantic elements representing concrete objects normally with relatively short life cycle, larger dimensionality and narrow application area. Thus the concept-oriented database model provides very convenient means for knowledge and data sharing where higher level concepts are responsible for storing common terms with longer life cycle while lower level concepts store information with the semantics defined via the lower level items.

Sharing data and knowledge in terms of concept-oriented model means sharing a coordinate system where its dimensions are represented by concepts and coordinates within these dimensions are represented by items. Such a common hierarchical multidimensional coordinate system or a set of common terms allows us to have a common interpretation of other item positions because their semantics is represented with respect to those terms. The concept-oriented database model does not make a distinction between shared terms and represented objects because each item is simultaneously an object represented by means of other items and a new term used to represent other objects. The selection of common terms or space structure in this case is reduced to choosing an appropriate level of details, i.e., a set of higher level concepts that will be used to represent all other objects.

Subconcept-superconcept relation between concepts is essentially equivalent to *is-a* relation, which says that any instance of more specific concept is also an instance of more general concept. Instances are included into their concepts by means of *instance-of* relationship.

An important aspect in representing data is that each item has a special unique value assigned to called reference. This value distinguishes an item from other items within one concept. Thus on one hand each item is characterized by its reference and on the other hand it is simultaneously characterized by a combination of other items. References are an absolutely necessary mechanism for data representation. Each reference is dual to the corresponding combination of item characteristics just like concept extent is dual to its intent. An existence of some entity that can be physically separated from an object yet retaining an implicit association with its characteristics is an important feature of any representation mechanism. In our case it is reference that is separated from an item and used to *represent* it in other items as an attribute value or characteristics.

Having references as a representation mechanism we can use them to store information about item characteristics. An alternative approach frequently used in theory uses no references or does not introduce them explicitly. In this case items are distinguished by their properties only or this mechanism is not specified at all. References provide not only a mechanism of representation but also allow us to connect identity and life cycle with an item reference rather than with its characteristics. In other words, an item exists and distinguishes from other items as long as its reference exists and is different from other references. In particular, we can change item characteristics and this will not change the item identity, i.e., it will still be the same item within the same concept. This is not true for other formal approaches and representation mechanisms. The existence of references is very important for the whole concept-oriented paradigm because they are one of the key elements of this approach.

## 2.4 Item Inclusion Relation

Concepts provide multidimensional hierarchical space where items live within structured containers. Each concept is characterized by a set of dimensions, which point to other concepts as domains for its values. Thus the structure of concepts describes database syntax or schema. On the other hand the items within these concepts describe the database semantics.

Each item is equivalent to a combination of other items specified as values for the concept dimensions. All items compose a structure where each of them has several parents and several sets of children (each set from one subconcept). We define each item as a combination of other items from the subconcepts. Each such combination is stored by reference in the corresponding items and the references are the primary means for representing information in the database. Notice that the reference structure is a directed acyclic graph so that an item may only have a property value from a superconcept.

However references play also a fundamental role from another point of view. In concept-oriented database model it is supposed that references implement inclusion relation rather than are merely a means for representing other items. It is important because inclusion relation among items is more important than its implementation. Item inclusion is of primary importance while there may be different methods of its implementation. Thus each item does not simply references a set of other items and thus specifies its relative coordinates. More important that each item specifies sets where it is a member.

Thus an important fundamental assumption is that specifying an item coordinate by storing a reference is equivalent to specifying a set this item belongs to. Thus each reference is interpreted as 'a member of' relationship among items. Since an item is a combination of other items it is included simultaneously in all its superitems. It should be noticed that normally we clearly separate references from interpretation of relationships among entities. References can be used without constraints as an implementation mechanism while logical relationships such as item inclusion must satisfy certain rules. This is particularly why many theoretical approaches do not consider references at all and pay main attention to some logical structure with no indication how it can be implemented. We follow a fundamentally different approach where references play a major role by supposing that any method or paradigm that is going to formalize or use the notion of representation and access must specify a mechanism by means of which elements know each other or store information about each other. It simply does not make sense to talk about an item with some relationships to another item if we do not specify how they know each other. In this sense the primary focus of the concept-oriented paradigm representation and access.

Such a principle can be viewed as a consequence or another side of a number of physical laws. We know that nothing exists by itself and no interactions are transferred instantly. In other words, for any physical relationship we need some mechanism or some implementation or underlying environment. What is even more important is that this mechanism is an integral part of the whole system and we cannot ignore it. If some object knows about another object then this knowledge is implemented by means of some other objects, which are also elements of the whole system and play an important role in its interactions.

Thus the principle 'reference as an inclusion relation' plays a primary role in the concept-oriented database model. In particular, this is why references in this model may have only one direction from subitems to superitems. It is a consequence of the fundamental inclusion relation among items. Notice that references also express 'references as attribute-value relation' described in section 2.3. In other data models such a reference constraint either does not exist or is not considered a primary principle.

References can be also viewed as 'knows' relation among items. If an item has a reference to another item then it knows it. This relationship plays a very important role in system design and can be used as a primary relation equivalent to the inclusion or coordinate relations all implemented by references. Indeed if a system consists of components then to establish a certain structure among them one of the things we need is knowledge about their existence and knowledge of their functions. This determines many other issues such as position in the layered architecture.

The 'knows' relationship is essentially an important aspect of inheritance relation in object-oriented programming. The thing is that an extension always knows its base object while base object never knows when and how it will be extended. Thus if an object knows another object then it is a form of extension because such an object can use functions and manipulate the known object. And vice versa if an object does not know other objects then it simply provides its functions and services for external use as a element of the basic level of the system.

Thus references among items are not simply an arbitrary mechanism but rather they provide the basic structure. In particular, references express the following types of relationships among items (Figure 4):

- Inclusion or membership relation where referenced item is a set
- Attribute-value or has-coordinate relation where referenced item is a value
- Knows relationship where referenced item is known
- Extension relation where referenced item is a base
- Characterization relation where the referenced item is a characteristic

Although all of them are in equivalent we will consider inclusion relation as a primary from theoretical point of view while other interpretations will be used to provide appropriate interpretations in different application areas.

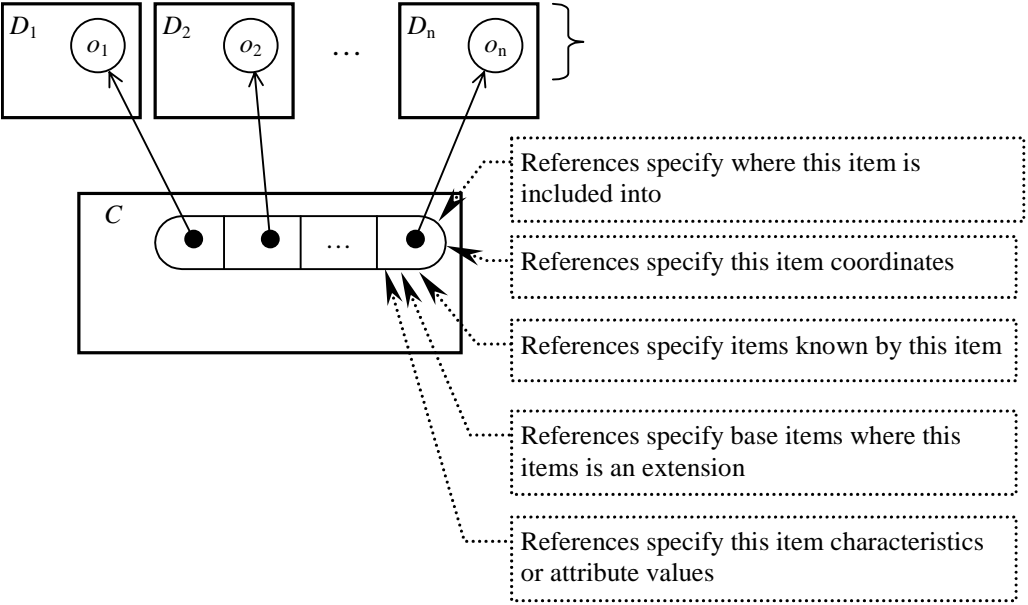

*Figure 4. Interpretation of references as various types of relations.*

## 2.5 Item Life Cycle

It is supposed that structure of space where objects exist is described by the syntax or schema of the database. After that this structure can be used to create items, which determine the database semantics or contents. Each new element is created within its concept where it is automatically provided a unique reference. (In practice references may have a custom structure and be generated manually.) Each new item has to be given its characteristics, which determine its position with respect to other items. In other words, we have to specify values for all attributes of this concept. For each such value its reference is stored in the combination of values corresponding to the new item. It the new item is not characterized by some variable then null value is assigned as this attribute value.

If an item is deleted then all its references in other items are assigned to null. This means that all items that have been characterized by deleted item have no value assigned to the corresponding attribute at all. We say that such items are not characterized by this attribute.

Some concepts may enforce having non-null values for some variables, i.e., nulls are prohibited in assumption that valid items may exist only with some value assigned to this property. In this case new items cannot be created with null values for these attributes.

What happens if such an attribute is nullified since its value has been deleted? Indeed, on one hand we cannot characterize it by the previous value because it has been deleted but on the other hand no new value can be automatically assigned and the null value is not allowed. In this case we suppose that since such item cannot exist with null value as its characteristic for this attribute it has to be deleted. Thus such deletion is a consequence of deletion of some superitem. In other words, propagation of nulls down to subitems may result in propagation of deletion operation. In this case deletion of an item may result in deletion of all or some of its subitems because they cannot exist with null values assigned to some their variables (Figure 5).

Null value propagation is a mechanism for deleting all subitems of non-existing item. In this case an item can be deleted if some of its superitems is deleted. However, in some cases we also need to delete an item if some its subitem is deleted. If the first mechanism reflects impossibility for an item to exist without its parent then the second mechanism reflects impossibility for an item to exist without being referenced by at least some child item. In other words, in some cases we want an item to exist only as long as it is used, i.e., there exists at least one subitem that references it. Accordingly as soon as all subitems are deleted such an item has to be also deleted. Such mechanism is called garbage collection since items are deleted when they are not reference anymore (Figure 5).

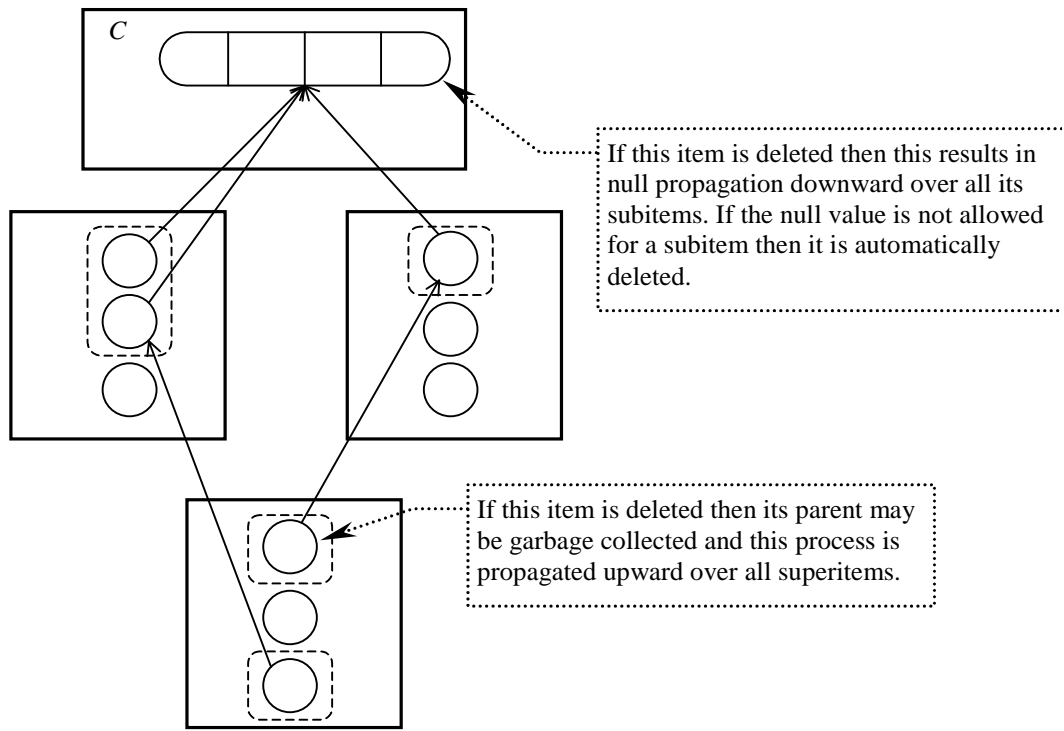

*Figure 5. Item deletion by null propagation and by usage control. If an item looses some of its necessary superitems then it is automatically deleted. If an item looses some of its necessary subitems then it is automatically deleted (garbage collected).*

In general case the garbage collection mechanism is applied to any concept, i.e., this mechanism looks after its item usage in other concepts and can delete items depending on the parameters set for this concept. The parameter to control the garbage collection of items is the scope for reference counting. In the simplest case we can simply turn this mechanism off and this effectively will make all items persistent or globally existing. We suppose that their scope extends this database where the garbage collection does not work so that the items can be deleted only explicitly. Essentially we say that somebody else has to look after the item usage. Normally this means that external applications know when items have to be deleted depending on existence of other objects (real or virtual). However, we can declare the concept local and restrict the usage of items by this very database so that the garbage collector can automatically delete them when they are not used.

To correctly assign the deletion options for concepts it is important to understand the logic of their life cycle management. In particular, it is necessary to find out if an item is able to exist without its superitems (null propagation mechanism) and if it can exist without its subitems (usage control by means of reference counting and garbage collection).

## 2.6 Levels of Representation

It is important that each concept has a certain position among other concepts and this position determines the corresponding level of details or scale used to represent objects. By level of details or scale we mean the effective number of dimensions used to represent items, which determine the number of features used to distinguish them. Concepts, which are positioned closer to the bottom concept, are more specific because their items are distinguished by all dimensions from superconcepts. In other words, the items from such specific concepts are highly dimensional so there exist a lot of features by which they can be distinguished. On the other hand items from concepts closer to the top concept are more general and are characterized by a little of dimensions.

In concept-oriented database model the database is considered as one whole, i.e., all items exist as an element of some object or a common part of a set of objects. The database knows about this stricture and its maintenance is its primary task. The selection of individual items, say, from one concept is an operation that simplifies our view of the whole database semantics. In other words, instead of viewing a complex set of items with their connections we can ignore some items and some connections so that the representation is getting more suitable for certain purpose. Yet the problem is what part of the whole representation we need to select in order to get an appropriate view. One approach consists in choosing some level of details.

For example, a country administrative structure is represented by describing all its administrative constituents such as states and counties (Figure 6). However, in many cases we need only a very general view of the country where it is considered consisting of only states while counties have to be ignored. For that purpose we might represent a set of all countries as one concept, a set of all states as a more specific concept and a set of all counties as even more specific concept. The states have a dimension specifying a country it belongs to. The counties have a dimension specifying a state they belong to. Thus we have three hierarchically ordered concepts where each country-item consists of a set of state-items, which in turn consists of a set of county-items.

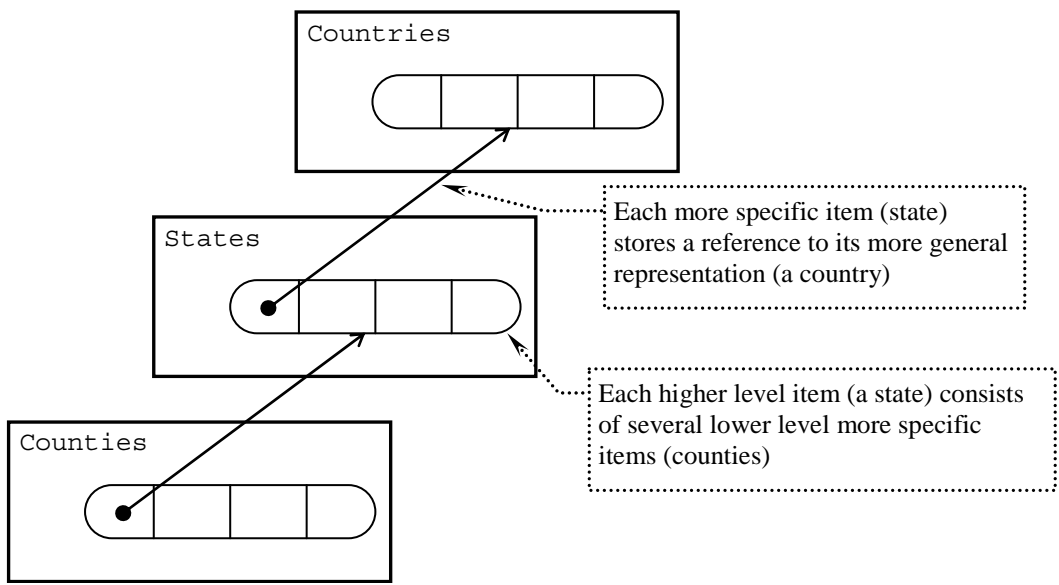

*Figure 6. Hierarchically ordered concepts produce different levels of details.*

At highest level the whole problem is represented by one item and we do not distinguish any details. Then we can decrease the level by choosing some lower level concept and then the problem domain will be represented by its items. Notice that here we normally have an alternative views represented by different subconcepts. This means that one and the same problem domain can be considered from different sides as different sets of objects. For example, a country can be considered as a set of its administrative units or as a set of companies. These are two dimensions that allow us to produce two different views on the country structure. We can continue this process and choose the next subconcept and thus produce even more detailed view of the problem domain where it consists of more specific objects. The most detailed representation of the problem domain is produced by specific concepts close to the bottom concept. Their number if normally high and they are characterized by many dimensions.

It is important that by choosing an appropriate level of details we can then bring the database semantics to this level by automatically propagating and aggregating information in all lower level more specific concepts.

### 2.7 Merging and Splitting Concepts

Normally a concept structure is designed using informal principles so that frequently we cannot say precisely why one structure is more suitable for the problem at hand then another. However, formally we can use operations to change this structure in such a way that the semantics is retained. Such operations are useful for both theoretical analysis and practical use, for instance, while implementing database schema refactoring methods.

One of such operations is concept merging (Figure 7). This operation consists in removing a superconcept and moving its syntax and semantics into the corresponding subconcept. The subconcept then gets all dimensions of its merged superconcept, which replace the corresponding dimension pointing to the superconcept. The superitems referenced from the subitems are moved by value and replaced their references. Notice that if there is more than one subconcept then the superconcept is repeated in each of them.

If this operation is applied recursively then the whole database can be represented as one bottom concept with very long items. The item length is equal to the database primitive dimensionality. In order to distinguish such merged dimensions we will denote them as paths, which starts from the subconcept name and then enumerates all intermediate dimensions. When all concepts are merged then in this canonical representation all such paths end with a dimension with primitive domain. Notice that items loose their identity after this operation because the merged superconcept is removed with all item references. After that the items from the superconcept are distinguished only by value.

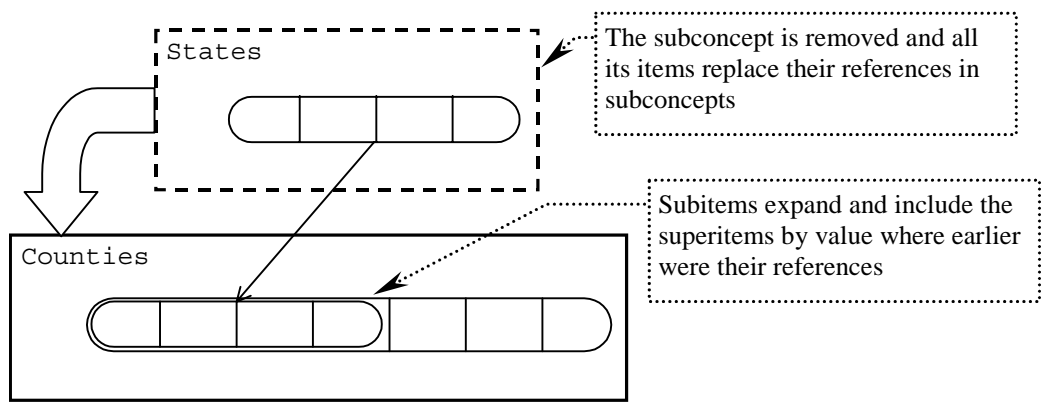

*Figure 7. Merging concepts. The superconcept is removed and syntax and semantics is moved to the subconcept.*

The opposite operation for merging is splitting concepts (Figure 8). This operation is applied to one existing concept and results in producing one its superconcept by selecting a subset of dimensions and moving them into this superconcept. The original subitems then store references to newly generated superitems from the superconcept. Notice that items moved into the superconcept obtain their own identity, i.e., their references generated by the superconcept.

This operation could be used to refactor the database schema in situations where new representation is more appropriate. For example, it could be applied to produce new set of items in the case they are going to be used in other subconcepts rather than only in the original concept. Or it can be used in the case we want to distinguish items by reference rather than by value.

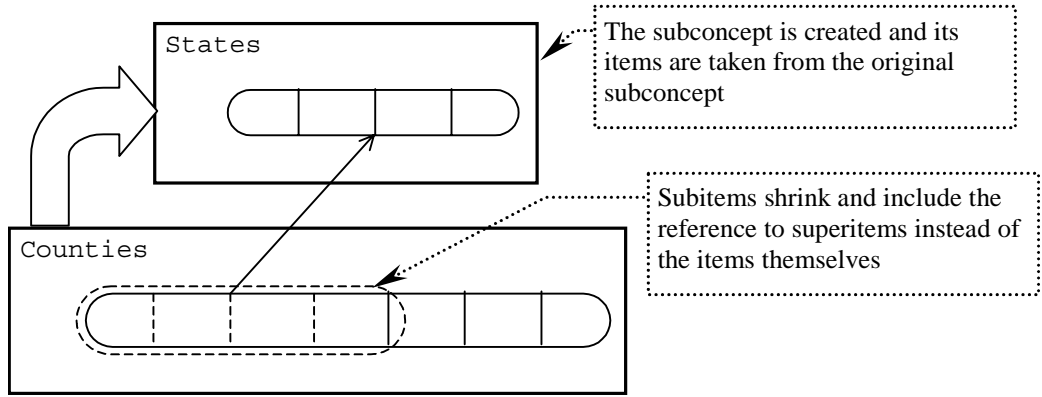

*Figure 8. Splitting concepts. A part of the original concept dimensions and the corresponding semantics is moved into the new superconcept.*

# 3 Query Language

## 3.1 Principles of Query Language

For any database model its query language should describe how a part of the whole semantics or new semantics generated from some part of the database can be obtained. The structure of the query language and its principles depend on the database model where it is used. For example, the relational model query language uses tables and columns as primary terms for its query language so essentially it is a language for manipulating tables.

An important specific feature of the concept-oriented model is that all elements exist in connection with other elements and it is the primary purpose of the database to maintain integrity of the item connections. In other words, the emphasis is made on connections rather than on items themselves and it is the database rather than the user who has to guarantee consistency of all connections. In this sense such a system might be called an item connection management system. Taking this into account any concept-oriented query language should provide facilities to manipulate connections among items. In particular, any item in the database has to be accessible from any other item following a system of connections.

The general paradigm we use for concept-oriented queries is that information from items in different concepts has to be expressed as properties of this concept. In other words, we suppose that there is always one concept the items of which we need to select. It can be one item, a subset of items or all items but we always know what kind of objects we need and choose them by specifying one target concept. The target concept can exist in the database or it can be virtual concept or view with items produced from other items following some rules.

Each concept is characterized by a set of dimensions, which can be considered its item properties. These properties in fact refer to other concepts so when we characterize the items we simply specify items from superconcepts. These properties can be used to restrict the number of items returned and we can select some of them to be returned as a result. Yet such a characterization by means of dimensions is very limited and in general case we would like to have a more expressive mechanism to collect information from all over the database associated in one or another sense with each target item. Such a mechanism has to be able to access information in all parts of the database and then transform it into a form this concept item property. The association is made possible by means of the database structure, which specifies how items in various concepts are connected.

Another aspect of any query language is that it can be viewed as a source program, which defines a sequence of operations that needs to be applied to data in order to produce some result. The specific feature of such operations is that they manipulate mostly elements within collections rather than individual variables. The problem is that simple queries are easier to define in a declarative form, e.g., by specifying the target concept and a set of dimensions while more complex queries can be expressed only using a style close to conventional programming languages. So there is always a tradeoff between simplicity of declarative query style and expressiveness of procedural query style. In particular, this is why the conventional SQL is not enough for complex queries and most databases provide additional languages for expressing more arbitrary requests to the database semantics. It should be noticed that in order to be executed any query is transformed from its source representation into a form, which can be executed by the database. This final representation is actually an internal program that is executed by the database virtual machine. In particular, any declarative representation is simply a short form for a small program, which is translated, executed and returns some result. The final transformed representation has one and the same form independent of the source query and it is a set of instructions to be executed by the database engine. In contrast to conventional procedural programs such instructions are designed to work with collections so a database query language could be called a collection-oriented approach. In particular, the collections are used as source elements and they are produced as intermediate and final results.

The concept-oriented query language described here is only one possible variant intended to demonstrate the main features of this approach. Therefore it can be convenient for one purpose but inconvenient or limited for other applications. In this sense one alternative might be to modify one of the existing query languages such as SQL. However, we describe our own version of the source query language in order to emphasize all distinguishing features of the new approach.

Since any query language is a source program we will use the formatting as accepted in programming, which is different from the previous sections where we used the formatting accepted in mathematics.

## 3.2 Item Selection

The central element of the described query language is a concept, which is a set of items. Thus we can say that the primary purpose of this query language consists in manipulating collections of items stored in concepts. The result of the query is a new concept, which is composed from some items stored in other concepts.

Each new or existing concept is identified by its name. By convention this name will start from capital letter, e.g., `Persons`, `Countries` or `C` are three concept names, which represent collections of items. Each database includes a set of existing concepts with items specified by value, which are analogous of conventional tables. Thus any query can use names of such concepts to produce new concept.

Concepts can be specified by value within the query by enumerating its elements, e.g., `{"Germany", "France", "Italy"}` is a new concept consisting of three country names and existing in the context of this query. Any concept can be given a name, which can be used later to access its items, e.g., `Countries = {"Germany", "France", "Italy"}` is a concept with new name that can be used further in the query to access its elements.

Items in collections might be also specified by reference rather than by value if item names are available: `Countries = {country1, country2, country3}`. If elements involve several dimensions then their values are specified in angle brackets: `Countries = {<"Germany", 80>, <"France", 50>, <"Italy", 40>}`. Notice that here dimensions names are not provided. If they are needed then they can be specified after the list of items in angle brackets: `Countries = {<"Germany", 80>, <"France", 50>, <"Italy", 40>}<CountryName, CountryPopulation>`.

In general case concepts are composed in much complex ways and the concrete procedure for their item composition will be specified within curly brackets: `Countries = {...}`.

Normally collections are specified by providing some procedure or mechanism for generating its items rather than explicitly enumerating them. The most widespread approach consists in iterating through another concept and building items included into the new concept. The iteration can be considered a loop on each step of which we have the current item as an instance variable. To describe such a procedure we need to indicate the source concept name and optionally the instance variable that will point to the current item. For example the concept `R={c:C}` will include all items from the source concept `C` since no other conditions have been specified. This procedure simply iterates through the whole concept C and includes all its items into the new concept R. Here c is the current item reference, which is actually not used and therefore we can omit it: R={C}.

Alternatively, to separate variable from collection we might use in or from keywords instead of column. For example, we might write R={c in C} or R={c from C}.

Frequently we need to iterate through a part of the whole source concept in order to select necessary items. For example, we could iterate through items of the source concept, which belong to some concrete item in one of the superconcepts. In this case in addition to the source concept name we have to specify its dimension name or a sequence of dimensions leading to the parent instance. For example, the concept R={c:C.x.y} or simply R={C.x.y} will include items from C, which are bound to some concrete instance in the parent concept $Dom(C.x.y)$. Thus the new concept R is a set of subitems of some parent item (Figure 9).

The parent item has to be set by the external context and the name of the instance variable that stores a reference to this item has to be reconstructed by means of some rules. As it will be seen later such rules are quite natural so that in most cases it is not necessary to specify the parent instance variable at all. However, in the case on uncertainty we can specify the name of this variable explicitly as follows: R={c:C.x.y=p} or R={C.x.y=p}. Here we suppose that variable p is a reference to some concrete instance from the concept $Dom(C.x.y)$. The equality means that we need to select only instances of C, which reference this concrete item rather than all items from C. The instance variable c now iterates only through this smaller set of items to be included into the result concept.

We might also specify a constant value as a parameter: R={c:C.x.y=0x0a0b0c0d}. Here the parameter value specifies record reference, which is guaranteed to have the same value as long as the corresponding item exists and therefore can be remembered and then specified in queries.

This is very simple and frequently used method to select only a subset of all items belonging to some parent item. It is very convenient but quite limited. A general way to describe items included into the target concept will be described in the next sections.

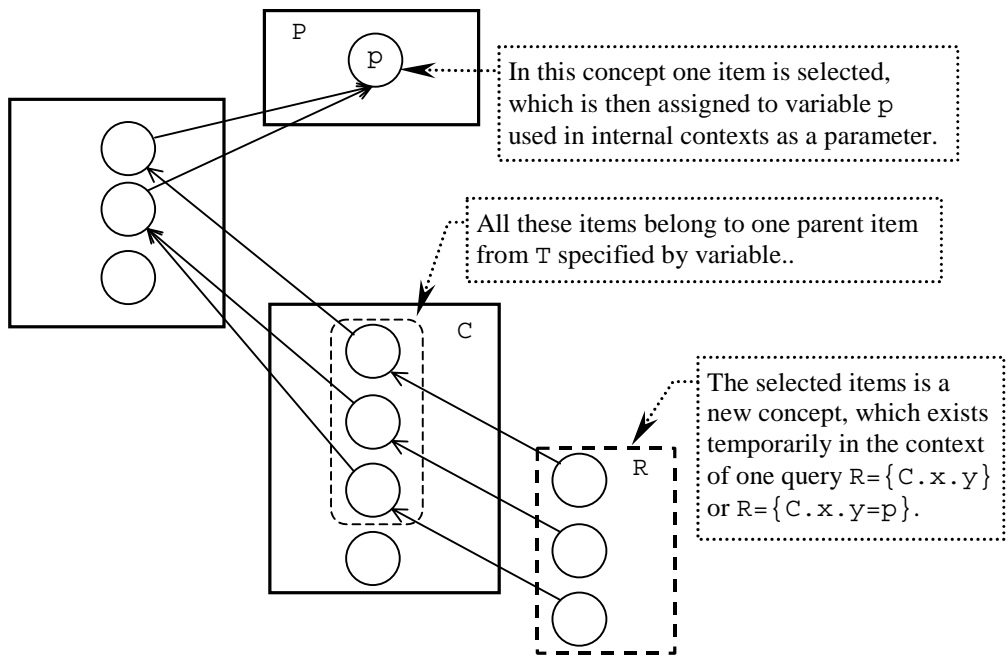

*Figure 9. Iterating through subitems of some concrete superitem.*

In some cases we would like to iterate through more than one existing concept simultaneously. For that purpose we can specify them in the query with the corresponding item instance variables: R={c1:C1, c2:C2}. In this case on each step of the loop two instance variables contain references to the items from the two source concepts. The order of iteration is not defined since it is the problem of optimization. Obviously such a query will iterate through all combinations of items from all the specified concepts and all these combinations will be included into the result concept.

The new concept produced by such a procedure stores only references to new items but no values of the original items. If we want to build the result concept, which include also some properties by value then the corresponding dimensions have to be explicitly specified. This can be done by enumerating the necessary properties after the concept definition: `R={c1:C1, c2:C2}<c1.x1, c2.x2>`. In this case the result concept R will include items consisting of two dimensions.

We can slightly generalize this approach by allowing for any formula to be used as a returned value. In other words, instead of specifying a name for the current item property we can provide a formula consisting of such names combined with various operations and/or functions. For example, the concept `R={c1:C1}<c1.x1+c1.x2>` will include items with the only dimension, which is equal to the sum of the two source item dimensions.

In SQL a set of source tables (concepts) is specified in the `FROM` clause with possible alias, which plays a role of instance variable. The columns included into the result are specified in the `SELECT` clause. For example, the SQL query `SELECT name FROM Country` is equivalent to the concept-oriented query `{c:Country}<c.name>`. Obviously here they are absolutely equivalent and there is no need for any special features at all. In SQL we can also use formulas in `SELECT` clause in order to return dynamically computed values.

### 3.3 Restricting Items

When a new result concept is being created from the source concepts then we normally want to include only a subset of all items in the result. There exists many methods for specifying what items have to be included into the result concept but the simplest approach consists in imposing logical constraints on item properties. Such constraints are specified as a set of elementary conditions combined by logical operations. For each step in the loop, i.e., for each item or combination of items this predicate is evaluated and if it returns true then a new item is created in the result concept. If the predicate is false then the loop continues by going to new source items.

The predicate restricting items to be included into the result concept is specified after the source concepts separated by bar: `T={c:C | predicate()}`. The predicate can be thought of as a function that evaluates the current source items and returning either `true` or `false`. Normally it is specified as a logical combination of elementary conditions imposed on source item properties. For example, the query `T={c:C | c.size > 10 AND c.weight < 5}` will return only items with the specified size and weight.

It should be noticed that in concept-oriented database model constraints are imposed only on the current item properties. This means that the decision is made always in the context of the current item, i.e., where the query variable is instantiated by the item reference. However, this does not mean that we can use only this item direct properties. Such an approach would be very limited. The thing is that the query language provides means for accessing other items in the database connected with this item and expressing them as the current item properties.

It is also possible to use values calculated by formula in restrictions. For example, the concept `T={c:C | c.population/c.area < 10}` will include only items with density less than 10. The returned values of the new concept are specified as usual after its definition.

Here we can see one drawback of such an approach (as well as SQL and similar approaches). If we want to use some calculated value and return it as a result then we have to write it twice: `T={c:C | c.population/c.area < 10}<c.population/c.area>`. Of course the compiler can always recognize such situations and compute the expression only once. More important that such a duplication breaks the structure of the query. In programming we always use intermediate variables where we can store the computed value to be used later. Suppose that such an expression is used within another expression, e.g., `T={c:C | c.population/c.area < 10}<10*c.population/c.area>`. In this case it becomes more difficult to reconstruct the correct computation sequence. This problem is solved by using the general query format described section 3.7.

## 3.4 Integrity Constraints

An important issue in any database is a mechanism for imposing constraints, which would restrict a set of possible states of the database semantics. In most cases this problem is reduced to constraining a set of possible data items in one table or concept. This means that for each concept we need to define what items are possible and what are not. Impossible items cannot be stored in this concept because they bring this concept and hence the whole database into the inconsistent state. Integrity constraints might be imposed by external applications however in most cases such an approach is not simply inconvenient but does not work at all. The reason for that is that most constraints must be checked within the scope of the current transaction, which updates the current state. In this sense it is important to select such a type of constraints and define them within the database while other constraints (non-transactional) could be checked by the corresponding applications.

The problem of imposing constraints is that the valid semantics of one concept normally depends on the current semantics of other concepts in the database. In other words, an item can be possible for a concept under one set of conditions and could be impossible for another set of conditions where conditions are defined by the current items in other concepts. According to the concept-oriented database model principles items do not have their own semantics at all, i.e., their semantics is defined via connections with other items. In this sense any constraints on possible items have to refer to other items in the database, which are somehow associated with this item. Thus the only way to impose integrity constraints on possible items is to check items in other concepts. The whole mechanism is then reduced to getting such an external semantics associated with items in some concept.

There exists many ways how the semantics from other concepts in the database can be brought to the selected concept, which are described later in the paper. However the general idea is that we formulate this semantics as concept properties. Once we have defined properties for a concept we can then impose constraints on their values and use logical operations to combine them. The properties can be simple or rather complex but for the mechanism integrity constraints their nature and origin does not matter. It is important only that given a concrete item its properties can be computed from the current database semantics and then checked against the conditions that have to be satisfied.

For each concept we can define a set of *inherited* (base) properties, which are taken from some its superconcept. In other words, each inherited property is equivalent to some path from this concept to some its superconcept consisting of a sequence of dimensions. Each such property returns a single value, which is either a reference to a superitem or the superitem itself if it is passed by value. Normally we are interested in getting primitive properties, which take values in the primitive concepts.

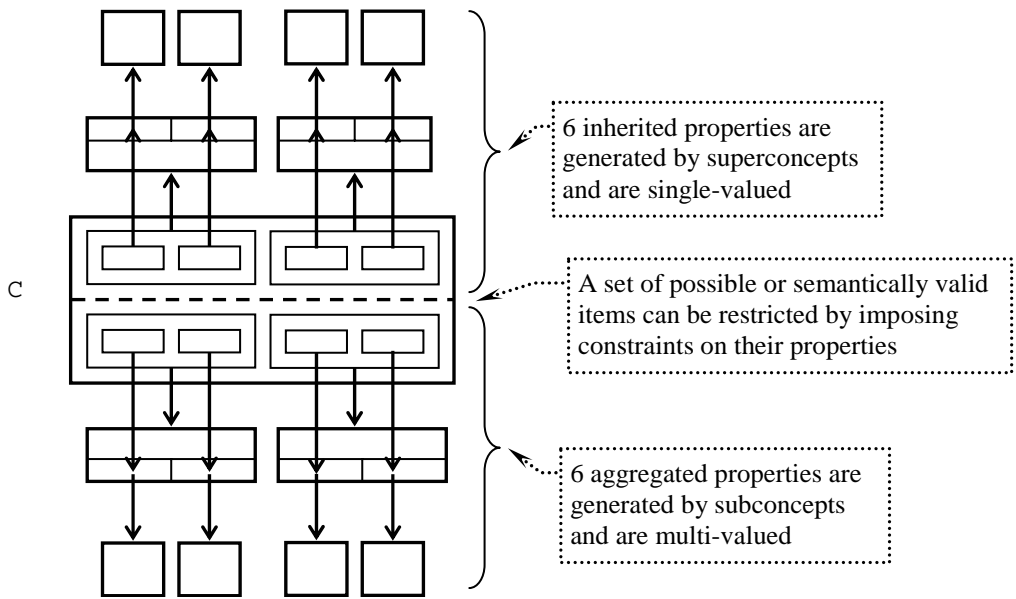

*Figure 10. Inherited and aggregated properties are used to impose integrity constraints on one concept.*

*Aggregated* (extended) property of a concept is taken from some of its subconcepts. These properties are multi-valued because their values are collections of subitems associated with the current item. More precisely each aggregated property returns a collection of subitems from one subconcept, which all reference the current item. In other words, all items in the collection have one and the same inherited property, which is equal to the current item.

In general case properties can be more complex because of the mixture of inherited and aggregated properties as well as applying various functions to them. Such complex properties are called *virtual* (computed) and are taken by following a sequence of links in any direction, i.e., it is a sequence of inherited and aggregated properties. Thus the virtual or computed properties may access information in any part of the database and the only important characteristic of this information is that it is associated with one item from the current concept.

All properties may have their names defined explicitly so that integrity constraints or other queries and access procedures could be defined easier. Once a set of properties has been defined the integrity constraints are imposed by restricting their values (Figure 10). For aggregated properties we frequently apply some aggregation function to the result collection.

## 3.5 Internal Concepts

Selecting items with imposed constraints is not the most difficult issue in any database. Much more interesting problem consists in accessing items from all over the database associated with other items. For this purpose we need to use the database structure or specify other criteria for associating items.

Each new concept defines a new scope where its instance variable takes some concrete value. It should be considered conventional scope in the sense of programming languages but it is instantiated on each step of the loop. Within this scope we can create and use other concepts. Normally such internal concepts are created depending on the state of variables in the external concepts, i.e., they are parameterised by instance variables set externally.

Such internal concepts are defined precisely as any other concept except that they use external variables as parameters restricting their items. If for external concepts we use constant parameters then for internal concepts we can in addition use values of variables set in their external context.

The most wide spread approach consists in using the current item reference to select its subitems. After that we can use this internal concept consisting of subitems as a property to decide if this item has to be included or not. For example, the concept R={p:P | size({C.x.y}) > 5} will choose only items from P with more than 5 subitems in the concept C linked by its properties x and y. More verbosely it can be written as follows: R={p:P | size({C.x.y=p}) > 5} or R={p:P | size({c:C | c.x.y==p }) > 5}. Here for each parent item p from P we create an anonymous internal concept consisting of subitems from C and then check its size.

Another example is where the internal concept includes not all subitems but rather only those satisfying certain condition. Moreover, this condition is not constant but also depends on the current context: R={p:P | p.z > sum({c : C.x.y | c.u > p.u }<u>)}. Here for item p in P we build its subsets from C but include into it items with property c.u larger that that in the parent item p.u. After that we check if the property p.z is larger than the sum of the selected subitems.

In these examples we applied aggregation functions such as size and sum to the internal temporary anonymous concepts. However it is not necessary and the intermediate concepts can be used in any other way just like collections in programming languages. For example, we might create an internal concept and then use it as a source concept to create another concept. The only problem is that the described form of queries is simple but quite limited to allow for such free concept manipulations.

The internal concepts can be nested. For example, the concept R={t:T | size({C1.x | size({C2.y}) > 5}) < 5} will contain elements from T, which have less than 5 subitems in C1. However in this set of subitems we include only the items that have more than 5 their own subitems from C2.

When we create internal concepts then normally include subitems of one parent item set in external context such as external concept definition. This parent item should not be necessarily the current item in the external context. We can choose also some its properties. As a result the path we follow to get information may change its direction and such a query is called zigzag query (Figure 11). For example, suppose that we need to select a set of items from T with the parent from P, which has less than 5 child items in C. We can write it as follows: `R={t:T | size({C.x=t.p}) < 5}`. Here the internal concept selects a set of items from C, which belong to one superitem from P, which is determined by the property `t.p` of the current item from T.

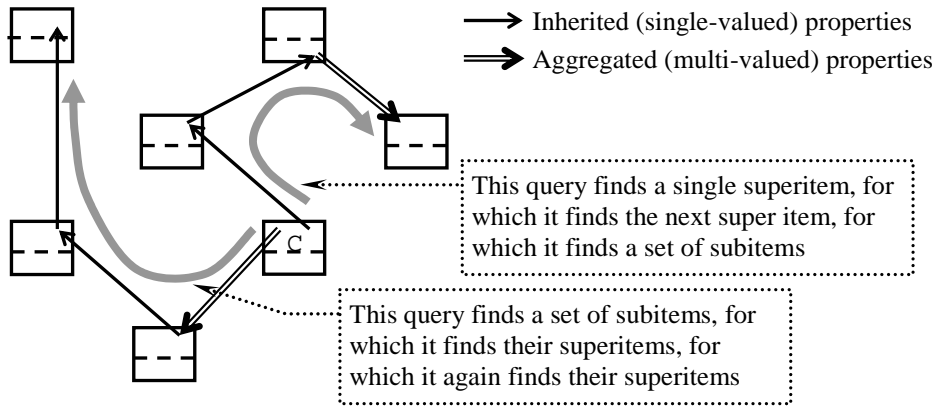

*Figure 11. Zigzag query. We can follow any path in the set of concepts following their inherited and aggregated properties. This path may even intersect one and the same concept including the original concept.*

We assume that the concepts computed internally within other concepts cannot be returned as their properties. This is done for simplicity in order to ensure primitive status of any item as a combination of other items but not concepts. In other words we want any item to include only other items (references or primitive values) but not other concepts. This is a very important fundamental principle of concept-oriented database model. All references have to point to a single super item while an ability of items to have sets of subitems is provided indirectly. In particular, an item cannot store a set of references or even one reference to its subitems. Such a possibility would contradict several major principles. For example, we suppose that subitems do not *know* about the existence of their subitems.

These principles are applied to concepts in the database and we also want to apply them to any concepts produced by queries. Thus we suppose that any new concept can return only references or primitive values but it cannot return internal concepts. Yet in practice this principle could be violated for the sake of flexibility and simplicity. In particular, a concrete database implementation might allow array types for dimensions or concepts-by-value type. In this case items are stored by value and this rather convenient approach for short lists. In the same manner we could return short internal concepts by value as dimensions of external concepts. For example, the query `R={p:P}<p.a, p.b, {C.x}>` will select all items from concept P and return its two normal dimensions and one dimension as a subset of items from the subconcept C.

Obviously the impossibility to return concepts as item dimension values is not a restrictive factor but it may simply make queries and result set processing more complex. Thus from theoretical point of view it is more convenient to have no returns of concepts while from practical point of view they could be allowed to simplify processing.

### 3.6 Property Names and Virtual Properties

An item has properties that are defined by its concept dimensions. The dimensions are defined at the concept creation by specifying a superconcept as a domain and some name. When a new concept is defined via query (a result set, a view or another kind of derived concept) we have to specify its dimensions just like for any other concept. In order to explicitly specify a new concept dimensions we have to provide its return values. For example, the concept `C={t:T}<t.a, t.b, a+b>` has three dimensions. However, these dimensions do not have explicitly specified names so the system has to choose them by default. In many cases it is easy to do, e.g., the first two properties in the above concept may have the names of their original properties a and b so that we can use these dimensions when creating new concepts. For example in the concept `R={c:C | c.a > c.b}` we use two these two dimensions to define new set of items.

However, there may be situations where the new dimension name should be provided explicitly. This frequently happens for computed properties. For example, we might want to provide a special name for the sum of two properties: `C={t:T}<t.a, t.b, sum=a+b>`.

In general case each new concept must have its own names for all its dimensions. The problem is only if these names are provided automatically or given explicitly. If a dimension name has to be given explicitly then we define it as a new variable with some definition. In general case this variable must also have an explicit type but here again we suppose that it can be omitted and then reconstructed by the system.

In general the main requirement to any concept-oriented query language is that it has to provide means for naming new concept dimensions. In other words, for any new concept created in the query we have to be able easily specify its dimensions in such a manner that after the creation this concept will be considered normal concept with normal dimensions.

The properties specified as returned values are persistent in the sense that their values are copied in the new concept. In many cases we would like to have virtual properties, which have a definition for their value but do not store the real value. Given an item we can get the value of such a virtual property but it has to be computed rather than retrieved from the storage behind the item. So this can be also called a computed property. Virtual or computed properties are defined by specifying a query, which computes this property value. After that the value of this property is computed each time it is requested.

There is also an intermediate variant where the value of property is computed and stored. However, it has to be then updated because the original values used to compute it might change. There may be different strategies and options for updating such properties.

## 3.7 General Query Format

In the previous sections we described a simplified query format, which is very convenient for more or less simple queries. However, in general case we cannot avoid its limitation. The problem is that an arbitrary access to the database semantics requires rather complex computations and in this case any query turns into a program, which has to be executed on the server. The distinguishing feature of such a program is that it manipulates concepts, i.e., collections of items. Thus just like for the simplified format such program is intended to compose a new concept from the source concepts. However, in contrast to the simplified query format the program provides much more flexibility and control over the whole process of selection of items.

The general query format includes several code blocks (Figure 12). These blocks are normal programs consisting of statements, variables, loop, conditions and other elements. However, in the query each block type is used in concrete moment of new concept composition.

First of all each new concept starts from a code block called `begin` and ends with a code block called `end`. The `begin` block executes before the main loop starts while the `end` block executes after the main loop finishes. For example, we might print a diagnostic message in these blocks or prepare some parameters or resource.

The other two code blocks are `before` and `after`. They are executed on each step of the loop over the items of the specified source concept. The `before` block is executed before the predicate is evaluated, i.e., for each item of the source concept. The `after` block is executed after the predicate is evaluated and only if the predicate is `true`. Thus the `after` block is executed for each item to be returned.

The next type of block is the `predicate` or the `from` block, which a function returning either `true` or `false`. This block is executed after the `before` block and if it returns true the `after` block is executed.

The `over` or `foreach` block specifies the source concept. In particular, it defines the instance variable and its range.

All program blocks are simply a sequence of code just like in any other program. These blocks are translated into a loop, which is executed on the server. These blocks can declare variables for intermediate values, which will be automatically destroyed after the loop finishes. Thus all variables created and computed during the query loop are transient and do not save their state. In order to return some result we have to indicate some variables as transient, i.e., the state of which will be stored persistently in the items of this concept.

```
begin();
forall(c:C) {
  before();
  if(!where()) continue;
  after();
}
end();
```

*Figure 12. A structure of general query. Each query consists of* `begin` *and* `end` *blocks,* `before` *and* `after` *blocks separated by* `predicate` *block, and attributes specifying return values.*

## 3.8 Database Views

Each concept is characterized by syntax, which determines its state space or a set of all possible points. This syntactic space is equal to the Cartesian product of all the domains, i.e., any combination of domain items is considered a possible point or state of the concept.

Notice that there exist two type of semantics: positive and negative. Positive semantics says what really exists or is true or has occurred. In this type of semantics it is supposed that we declare what is necessary (exists, true, occurred etc.) while everything else that has not been declared is unknown. For example, if we say that the point consisting of the combination of values <"John", 30, "Michigan"> is true then this means that we declare our positive knowledge about this fact but we still do not know anything about other combination. In particular, we do not know anything about the combination <"John", 50, "Michigan">.

The negative semantics declares what is known to be impossible, non-existing, or that cannot not occur. This type of semantics is normally used to restrict the set of possible points, which is declared syntactically. For example, we might declare that the combination <"John", 200, "Michigan"> is impossible (assuming that 200 is age).

There is two major way to define concept semantics of both types. The first approach consists in assigning each individual possible point one of two states or semantic characteristics. These semantic characteristics may have different interpretations depending on the formal settings, e.g., as 'point real existence', 'point is true', 'point has occurred' etc. Generally the semantic space, i.e., the number of characteristics could be larger, e.g., points could be assigned values from the interval [0,1] interpreted as probabilities but database theory normally assumes two-valued semantics while all other cases can be simulated. Such an approach is extensional in the sense that we can only manipulate semantics in one point. We can declare a point as positive (true, existing, necessary etc.) or negative (false, non-existing, impossible etc.) however it is only one individual point that can be referred in one semantic statement.

An alternative approach consists in assigning a whole interval of points some value. For example, we might say that all combination of values with one specified dimension greater than 200 are impossible.

There is also the third approach, which provides a procedure or algorithm or a set of rules for generating semantics from other available information. It looks similar to the intensional method where semantics is represented implicitly. However, the difference is that the semantics in this case is not independent and is produced from information, which already exists. Thus the idea of this approach is to change the form of the semantics. The form is changes by some procedure or transformation but the final semantics is stored in the conventional concept (really or virtually). In other words, given a concept we are not able to determine if its semantics is independent or derived from some other semantics because by retrieving information we can only determine if some point is true or false.

Views are concepts with semantics derived from other concepts by defining the corresponding procedure. For views we have to specify their dimensions as usual but the semantics is defined differently. Instead of assigning point some semantics value explicitly we provide a procedure, which computes such a semantic characteristic from some other data in this database. There exist different mechanism how we can define this procedure. One approach is based on logic programming where the view is provided a predicate, which then computes for each point if it is true or false. Another approach uses procedural programming, where for each point the procedure computes if it is true or false. The third method consists in computing the list of all true (or false) points. In any case it is important that we are not able to insert, delete or update items in views.

The most wide spread approach for defining views consists in using the corresponding query language. In the concept-oriented model we also follow this approach and hence each view has to be defined by specifying a query, which returns some collection of items. This collection then is considered the concept semantics.

# 4 Data Modelling and Analysis

## 4.1 Principles of Schema Design

One of the main principles of concept-oriented database model is that each item stores a combination of references to items from other concepts and all the references must constitute a directed acyclic graph with no cycles. Such a structure is guaranteed by the corresponding structure of concepts where superconcept-subconcept relationship must also constitute a directed acyclic graph.

In order to create such a data model for a concrete problem we need to follow some principles for interpreting relationship among concepts and their items. In particular, the following interpretations of the relationship among concepts and their items can be used:

- Inclusion or membership relation, which means that items from a subconcept are elements of items from its superconcepts or vice versa, a superitem is a set with elements as its subitems. For example, an organisation item consists of departments and a product consists of its parts.

- Attribute-value or has-coordinate relation, which means that a subitem has a unique position in some domain identified by its superitems or vice versa, superitems denote some positions in their domains to be taken by the subitem.

- Characterization relation, which means that each subitem uses its superitems to characterize itself and determine its position among other items. For example, a product could characterized by a producer while a producer could be characterized by its country of origin and its address.

- Knows relationship where subconcept and its items know their superconcepts while the superconcept never knows what possible subconcepts may already exist or appear in the future. For example, product parts never know existing or future product where they are used. Just like products never know their customers and order number used to buy them.

- Extension relation where subconcepts add some new dimensions to their corresponding superconcepts while superconcepts represent base dimensions common to a set of other concept extensions.

When building a concept-oriented database schema for each already defined concept we should ask the following questions in order to determine its possible subconcepts (top-down approach):

- What sorts of items each item from this concept includes as its members, i.e., what are other items each item from this concept consists of? For example, for country concept we might say that each country consists of its regions and from its companies while for company concept we could decide that it consists of its departments and products it produces. For each such sort of items we should create a subconcept of this concept.

- What are other sorts of items for which this concept serves as a domain or dimension, i.e., what are other sorts of items for which these items are coordinates or attribute values?

- What other items are characterized by items from this concept?

- What other items know items from this concept?

- What sorts of more specific items extending items from this concept exist? Such extending concepts add new dimensions to their base concepts.

Once we answer these questions we can add more subconcepts in the schema and after that this process is continued so that each new concept uses already existing concepts to define its structure.

However the situation normally is not so simple. The main difficulty is that some concepts are not formally comparable in the sense of the subconcept-superconcept relation and accordingly their items are not comparable in the sense of subitem-superitem relation. For example, if we have already the concept of products, which consists of concrete product items then it seems quite natural to introduce its concept of companies so that each product has a set of subitem companies, which produce it. However, here can easily notice that each company may produce several products. Obviously it is precisely the case of many-to-many relationship because one product is produce by many companies and one company may produce many products. How can we describe this relationship in concept-oriented database schema? For this purpose we have to define an auxiliary concept, which is a common subconcept for both products and companies. This subconcept will contain items, which characterized by a single product and a single company (and may be some other properties). Thus it is important to understand that when entering new subconcept with items characterized by some existing superconcepts we have always to check whether this relation is unidirectional, i.e., these concepts are comparable. Otherwise it is necessary to introduce some auxiliary concept. Notice that in real-life databases the number of such auxiliary concepts storing instances of complex relationships can be rather high. The concept-oriented database model is very convenient for modelling such cases because of its hierarchical structure. In other approaches, in particular, in relational model it may cause serious difficulties because all tables are equivalent and have the same level.

On the other hand if we have already a concept then we need to determine its dimensions (bottom-up approach). In the concept-oriented database model all dimensions are identified by their domains, which is simply another concept. Thus for each concept we must indicate a combination of other concepts to define its structure. To determine an existing concept dimensions we can ask the same questions:

- In what other items these items are included? What groups constitute these items and these groups are represented by other items?
- What are the characteristics the items from this concept take?
- What other items characterize the items from this concept?
- What other items are known by the items from this concept?
- What other items can be considered the base part of the items from this concept?

## 4.2 Relationships

The conventional approach to data modelling consists in defining two types of elements: entities and relationships (in fact, this approach is much more general and is applied in different form to many other areas). Entities are associated with objects while relationships are used to establish connections among them. Its most wide spread form this approach found in entity-relationship model. Although this approach is very natural and is widely used in practice it has some serious drawbacks. The most serious problem is that it is extremely difficult to distinguish between entities and relationships because relationships are in most cases also entities. So essentially there is no theoretically substantiated method to say if something is an entity or relationship.

Thus we come to one of the cornerstones of the whole concept-oriented paradigm: everything is an object including relationships among them. It looks like a contradiction but it is precisely what is assumed in the concept-oriented approach. We simply suppose that objects may have different levels and hence play different roles with respect to objects on other levels. In other words, we interpret an object as an entity with respect to objects on one level and interpret this very object as a (instance of) relation with respect to objects on the other level. This role is thus relative and cannot be assigned unambiguously.

Although objects (items, entities) exist on different levels and can be interpreted as relationships they still need some mechanism to establish links. For this purpose we use references, which are the only possible type of relations among objects. All the objects are connected by means of one basic relation called reference relation. This relation is used to produce levels and then objects from different levels can be interpreted as instances of relations.

The references look like an exception to the assumption that we deal only with objects, because they are definitely not objects. Indeed, we are not able to create an instance of reference and the only thing we can do is to establish a link from one object to another by means of the target object reference. References do not have properties and cannot be manipulated as objects. However, it is only an illusions and very convenient abstraction. In fact references are also objects but existing at lower level, which we do not consider.

The data modelling in concept-oriented approach is based on describing only one type of relationships between items and their concepts. This type of relationship is based on reference structure and may have different interpretations suitable for different applications, e.g., inclusion or knows relationship. Yet this does not means that we cannot use higher level relationships from conventional modelling techniques. As we mentioned any item in concept-oriented database model can be considered as an instance of the corresponding relation, which connects its superitems. Thus we can define concept structure and then interpret some its items as entities while other items will be viewed as relationships. Or alternatively we can start from defining what we mean by entities and what we mean by relationships by describing an entity-relationship diagram and after that this diagram can be transformed into the corresponding concept-structure.

The main idea consists in interpreting an item as an instance of relationship between its superitems and vice versa any relationship is translated into an instance of its subconcept (Figure 13). Thus any relationship is translated into a concept and concepts can be considered relationships. If we are given an entity-relationship diagram as a graph where entities are nodes and relationships are edges then it can be translated into the corresponding concept-oriented database schema. This schema has two levels. The first one consists of entity concepts while the second more specific level consists of their relationships subconcepts. Notice that according to the concept-oriented principle entities never know the relationships they are involved into while relationships know the objects they connect.

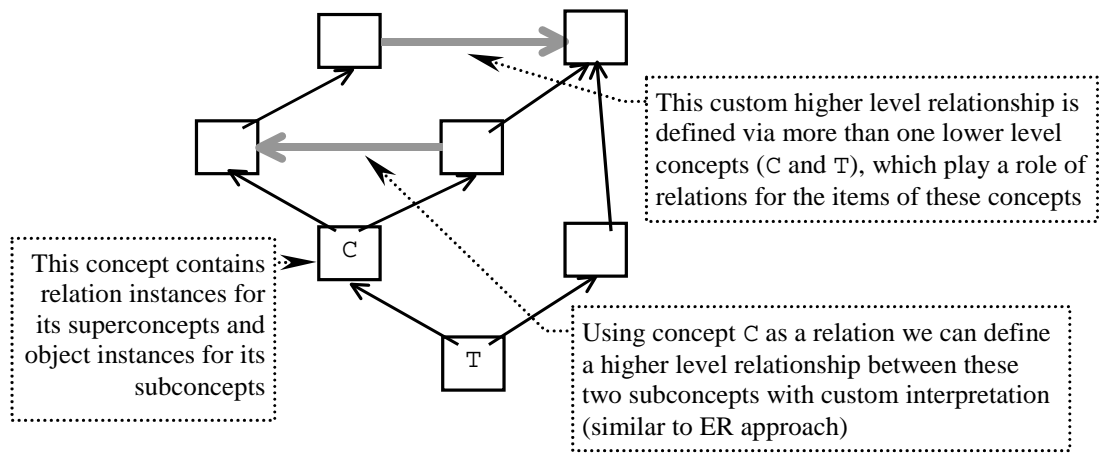

*Figure 13. Concepts are interpreted as relations for their superconcepts and as sets of objects for their subconcepts. Higher level custom (parallel) relationships between arbitrary concepts are defined via lower level concepts interpreted as relations.*

The entity-relationship approach is not so flexible because it assigns a concrete role to each element either as an entity or as a relationship while in complex applications such a role is always relative, i.e., a concept can be used both as an entity and as a relationship. Another problem is that in complex data models relationships may have rather high cardinality, i.e., their instances connect several superitems rather than only two. Such concepts are difficult to interpret and they may even have several interpretations as a relationship. The next serious problem is that it is very difficult to model hierarchical relationships. Suppose for example that a new relationship has to connect an entity and an already existing relationship or two other relationships. In diagram we would have to create a connection between two existing edges. Obviously such an approach is not acceptable while in concept-oriented model this is done very naturally for hierarchies of any depth.

Although the concept-oriented approach is more general the notion of relationship is also very useful in many situations. The main advantage is that we can use some concepts as relationships between other concepts. Relationships in concept-oriented model are abstract references or links between arbitrary concepts, which are implemented via more specific concepts. For example, if two concepts A and B have a common subconcept R then might directly use a query language or we might introduce an explicit relationship, which uses concept R to access items from B given an item from A.

Notice that such concept-oriented relationships are not primitive and have some implementation, which can be rather complex involving several concepts on different levels. For each such relationship it is necessary to provide its definition and generally there may be more than one relationship defined for the same concepts. Concept-oriented relationships are implemented via virtual (computed) properties. If we need to declare a relationship for a concept then it is done by defining a property, which is essentially a query, which returns a result concept or one item from the target concept. Thus each relationship connects one source concept with one target concept by returning one or more target items. Let us notice once again that the main advantage of relationships is that they establish arbitrary connections between concepts rather than direct reference-based connections.

It is important that in contrast to relational model in the concept-oriented database model there is no one-to-one, one-to-many and many-to-many relationships. There exists only one type of basic type of relation, which connect a subconcept with its superconcepts and this relation is single-valued by definition. All other relationships are derived from this basic relation.

For example, if there are two concepts Companies and Persons then many methods allow the modeller to introduce a relationship between them, say, "works in", which means that a person can work for a company. In the concept-oriented model we intentionally exclude such a mechanism because it has a secondary nature. Instead we say that relationships are simply views on the basic semantics. This means that if we want to have such a higher level relationship then we need to define it via basic elements. Moreover the definitions are ambiguous and there may be different types of relationships and different interpretations for the same basic semantics. Another approach is to define the basic structure and then interpret some of its parts as higher level relationships.

Another feature of the concept-oriented relationships is that they do not have names or identity. In such models as ontologies or ERM, relations have the primary importance equal to objects they connect. In this case relations are equivalent to dimensions while in concept-oriented approach we separate dimensions from relations as two different types of basic elements of the model.

The concept-oriented approach to interpreting relations essentially means that there are no relations at all — relation instances are concept instances of lower level. If we have a relation then it follows that we have to have some concept for that purpose. In our model there is only one sort of things while other things including relationships are derived from this primary type using some assumptions and adequate interpretations.

## 4.3 Multi-Valued Properties

The concept-oriented database model is based on reference relation, which has several interpretations but in any case an item stores a combination of references to items from other concepts. It is rather natural especially if interpret this relation as this item attribute values. However, in many situations we need to have attributes that take a subset of values rather than a single value. For example, a customer could be assigned a set of orders, an order is characterized by a set of products, and a person might have multiple addresses.

One possible solution consists in simply generalizing single-valued properties on the case of multi-valued or set-valued properties. This means that we make it possible for dimensions to take a subset of items from their domain as a value instead of only one item. In other words, if earlier an item was characterized by only one item from each domain then now items could be characterized by a set of items from some domains.

This approach appears rather natural solution to the problem and currently it is used in almost all data modelling methods. The direct consequence of such a freedom is impossibility to introduce a formal structure and semantics into the model what is however compensated by its convenience in practical use. Indeed whenever we need we can simply declare a dimension as a multi-valued with no restrictions (just like there is complete freedom with no restrictions in reference structure).

In concept-oriented approach such an approach is not allowed because it contradicts to our fundamental principles. In other words, in concept-oriented database model items can be characterized by only a single item taken from each its domain. There is no possibility to characterize any item by more than one another item and it is a high level principle determining the behaviour of the whole representation mechanism. For example, we cannot define a dimension called Orders for a concept

Customers so that each customer item will be characterized by a set of its orders. Thus strictly speaking the concept-oriented database model does not allow creating multi-valued properties.

Yet the problem exists and it has to be somehow solved, i.e., we need to have some mechanism to describe an ability of items to be characterized by sets of other items rather than only a single item. The concept-oriented solution to this problem is that items can be characterized by subsets of other items indirectly by interpreting references in reverse direction. In other words, groups of subitems where each group is taken from one subconcept are interpreted as this item multi-valued property or characteristic. Thus items do not store any references on their multi-valued characteristics and cannot get them directly.

In general case multi-valued properties are interpreted as arbitrary relationships of this concept with other concepts, which return subsets of items from other concepts with the help of the corresponding query definition. If multiple values are taken from direct subconcept then the query and interpretation is very simple. However, if a set of items somehow associated with this item and characterizing it from some side is taken from arbitrary concept in the schema then the query can be rather complex.

Manipulating multi-valued property is a very important part of data modelling. Although this problem can be solved in general form by means of relationships and standard queries in many cases we could simplify it by introducing auxiliary mechanisms. For example, in a data modelling tool we might want introduce an explicit notion of multi-valued dimensions, which can be added to any concept. Since the concept-oriented model does not support them directly they need to be implemented indirectly by means of the corresponding relationships.

For each new multi-valued property of this concept we need to specify its target concept, which can be anything in the current schema. After that the property can be implemented automatically by introducing hidden common subconcept with two dimensions combining one item from the source concept and one item from the target concept. Then simple query will return a subset of target items associated with one source item. Normally such properties are given names in order to simplify their use. In this case they can be used just like normal single-valued item properties, i.e., given an item we can write its multi-valued property, which will return a subset of its values.

The manipulation multi-valued properties could be simplified even further. For that purpose we can interpret them as concepts associated with some item. In other words, if for single valued properties we can only assign a value (a single item) or check the current value then for multi-valued properties the spectrum of operations is different because they are collections. In particular, we can add a new target item to the multi-valued property or delete an item from this property using the corresponding operations: `i.m.add(t)`, `i.m.delete(t)`. To apply such an operation we have to specify an item, then its multi-valued property name and then the operation name. We can also apply aggregation operation or check some conditions: `size(i.m)`, `t in i.m`. It is possible to define new queries based on one item multi-valued property as a source concept: `R={t:i.m | t.a > 5}`.

## 4.4 Grouping and Aggregation

One of the most important and typical tasks in data analysis is aggregation of information stored in many items. The problem is that information might be stored in low level items in too fine grained form while we need to have a higher level view where it is represented in a coarse grained form.

One approach to this problem consists in selecting groups of items on the basis of some criterion and then processing items in each group. The processing is normally applying some aggregation function such as the sum of values or average value. Then the main question is how groups are selected and what are the criteria used for that purpose.

In relational databases the grouping is based on the ability of items to have one and the same property. In other words, we first consider a set of all items that are characterized by different properties and then each group of items includes those with the same value of the specified attribute. This approach is very simple and flexible because does not impose restrictions on group formation. However it lacks formal semantics because groups have no interpretation in terms of other data items in the database.

SQL has a special clause for grouping records, which is called `GROUP BY` and is followed by some column name. For example, if we want to group people on their sex then we write it as follows: `SELECT * FROM People GROUP BY sex`. This statement should return two groups of records where each group consists of all people of one sex, i.e., one group of men and one group of women. In fact this statement will not work because we did not specify what the database has to do with information in those groups, i.e., what kind of aggregation function has to be applied to group members. The following correct statement will return the mean age of men and women: SELECT avg(age) FROM People GROUP BY sex.

The grouping criterion can be more complex and involve derived (computed) properties as well as the aggregation function can be any formula that is applied to all records from each group. For example, in data analysis we might be interested in aggregating information about groups consisting of persons with the same difference between their own birthday and their first child birthday.

In concept-oriented database model there is no separate mechanism for grouping and aggregation because this model has an intrinsic support of item association. In other words, in the concept-oriented model all items are already grouped in a natural manner corresponding to the problem domain. (Such an embedded grouping and aggregation mechanism is one of the main design goals of this model.) In this sense it is similar to OLAP where information is also aggregated using the built-in hierarchical structure.

In order to successfully use the grouping and aggregation facilities of this approach each item has to be interpreted as one group. Thus groups are primary entities and exist explicitly. In contrast, in the relational approach the members of groups are primary things while the groups themselves have a secondary order and are implicit objects. Indeed in concept-oriented model each item can be considered a group for its child items and hence we can always aggregate information in child items around their parents. Since each item is characterized by a number of parent items its information can be aggregated upward in several directions. Notice that parent items formally exist even for primitive domains.

In order to start grouping and aggregation we need to specify a concept, which contains the groups as its items. After the grouping concept has been determined we can use its multi-valued properties or arbitrary query to get subsets of items associated with each item. Each such subset of items from other concepts will be considered as a group associated with the corresponding primary item. Once such subsets of items have been defined we can apply any aggregation function to get the final (normally scalar) characteristic of the group. Notice that there may be different subsets selected for each group and different aggregation functions applied to each group (Figure 14).

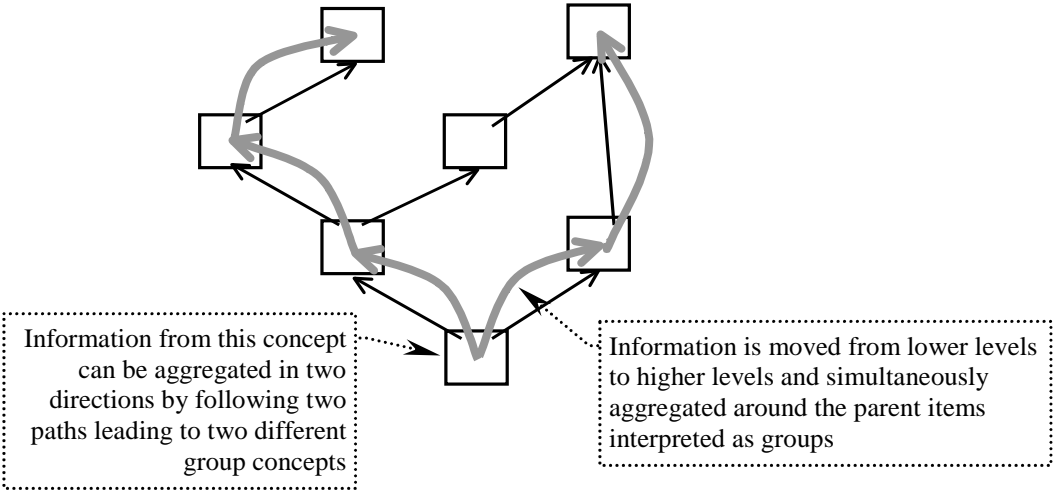

*Figure 14. Aggregation is based on the syntactic structure. Information is brought from lower more specific levels to higher more general levels. The groups are created around parent items.*

If a concept has multi-valued properties defined then the aggregation is especially simple. We need to select the group items and then apply an aggregation function to this property. For example, if m is a multi-valued property of item i then such an aggregating query can be written as follows: R={g:G}<sum(g.m)>. Here the result concept consists of items of the concept G, which are groups and for each this item its multi-valued property is computed and summed up.

If the concept consisting of groups does not explicitly exist then it has to be defined via query and then its items has to be used as usual as groups for which multi-valued properties are computed and aggregated. For example: the concept `R={g:G}<sum({t:T.g}<t.a>)>` will include all items from `G` (groups) for each of which we find its subset of items from subconcept `T` and then sum up its property `a`.

Frequently used aggregating queries can be defined as virtual (computed) properties of a concept. For example, we might define the balance of an account or the mean salary as such properties. The advantage is that such a property has its own name and can be used as normal single-valued property. In particular, such aggregated properties can be used to restrict group items depending on their aggregated information. In SQL such a filtering is made possible by `HAVING` clause, which is used in combination with `GROUP BY` clause.

Generally information can be aggregated in much more complex ways. However the main idea of aggregation is that it is brought up from specific level to more general ones. On its way from lower to higher levels the information looses its details and gets coarsely grained form.

### 4.5 Online Analytical Processing

The idea of online-analytical processing consists in aggregating information grouped on several dimensions and on different levels of detail. The dimensions chosen for item aggregation have a hierarchical structure and each combination of values (cell) of this space is used as a group. For each group or cell we can select a subset of data items and then aggregate some its parameter called a measure.

It is important that OLAP technology supports roll up and drill down operations to change the level of details for its dimensions. Thus this data representation and aggregation method does not provide a static view on data semantics but also allows dynamically change the level of details. Notice that such a variation of the level of details does not change the data semantics, i.e., we view the same data with the same characteristics. This is rather deep topic concerning the nature of object representation because here we come to the conclusion that objects spread their representation among many levels and our data representation hierarchy is only a small portion of the global space where objects live. Such a multi-level object representation is one of the key principles in the concept-oriented approach where one and the same object can be considered at various levels starting from high level logical representation and then spreading down deep into the physical interactions. The concept-oriented database model relies on this principle by defining each item as having several parent items and constituting from several sets of child items.

In order to generalize the representation and decrease the number of object in some dimensions (and the number of groups/cells in the OLAP cube) we apply roll up operation. On the other hand, in order to go deeper into details we drill down and add more elements into the representation so that the number of cells increases. In particular, we can reach the maximal level of details where all available objects are selected so that no aggregation is performed for them. Such a variation of scale can be viewed as change of the coordinate system. Since all items are identified by their coordinates we can vary this system by mapping fine grained coordinates into coarse grained coordinates.

The multidimensional hierarchical representation can be viewed as a generalization of the conventional (flat) multidimensional approach and the conventional (one-dimensional) hierarchical approaches. If there is one dimension then it is represented as one concept. Then we can introduce its superconcepts with less details and subconcepts with more details. Each such concept is still the same dimensions but viewed at different level of details or scale. For example, time dimensions could be considered consisting of years, quarters, months, weeks or days, which are ordered hierarchically but it is still time, i.e., objects will be characterized by different items in this dimension. However it is important that each time moment has to be linked to the corresponding time moments at lower and higher levels, i.e., all months in one year belong to one item representing this year. Such a structure can be viewed as a classification or categorization because each object belongs to one class represented by an item at some higher level (Figure 15). The class then can substitute its objects if we need to consider the problem at higher level of abstraction (lower level of details).

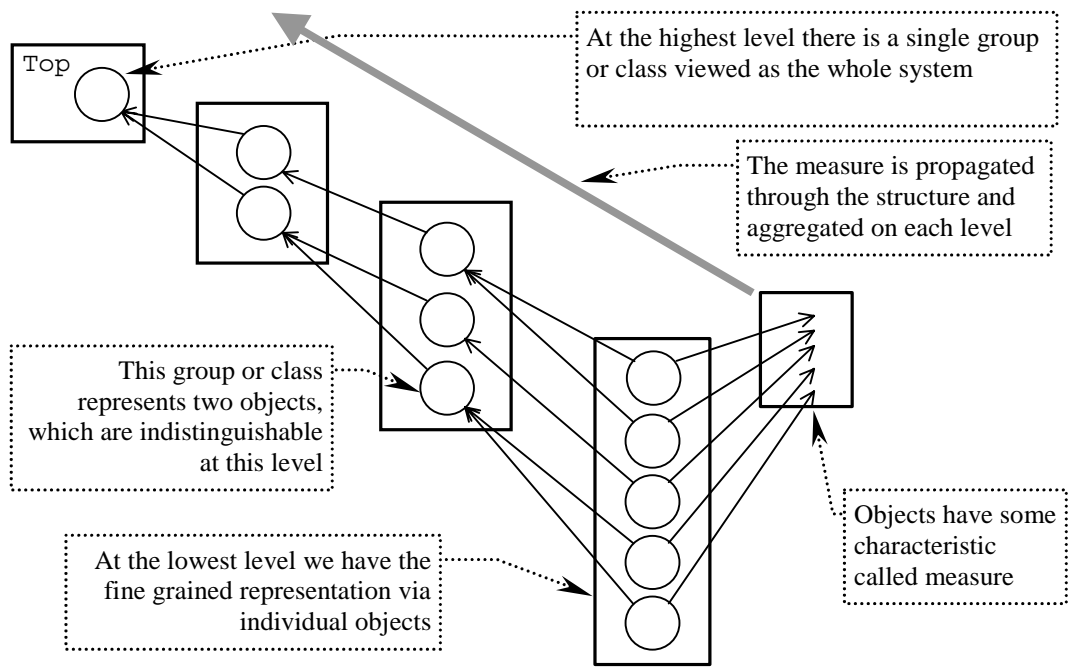

*Figure 15. Hierarchical concepts correspond to different levels of detail in the coordinate system. Such a system can be used for hierarchical classification or categorization. In OLAP terms such a hierarchy is called dimensions with different levels, e.g., it might be a product dimension with different level corresponding to product categories/groups/classes. The items of this hierarchy are called in OLAP dimension members.*

If we use only one classification structure then the problem is hierarchical but one-dimensional. We can easily add more dimensions by defining the corresponding hierarchies. For example, we might have a hierarchy of time, product categories, organisation hierarchy and so on. After that each object is assigned a coordinate along each of these dimensions so that it is correctly positioned in this multidimensional hierarchical space.

When all objects are positioned we can propagate this information in the upward direction by aggregating some properties called measures (Figure 16). Theoretically there is no distinction between dimensions and measures because they are both represented by superconcepts. However measures are typically of numeric type so that we can easily apply numeric aggregation functions. In other words, objects at the lowest level are characterized by dimensions and measures in the same manner. For example, one sale is an item, which is characterized by time, product identifier and amount. However normally we would like to aggregate the amount along other characteristics such as product categories. The aggregation along amount of sales also is possible but is used only in rare situations for more complex analysis. In any case it should be noticed that dimensions and measures are symmetric in the concept-oriented schema and their role is specified only at the level of aggregating queries and functions.

In concept-oriented terms the choice of one dimension consists in specifying a concept the items of which will be then considered values along this dimension. Normally several dimensions are selected for analysis, which are interpreted as axes of the OLAP data cube.

All the chosen concepts must have a common subconcept. This can be either a real concept existing in the database or a query. For each combination of dimension values, i.e., for each set of items from the dimension domains we can select a group of items from this common subconcept. These items constitute one group, which can be further aggregated.

In order to aggregate items in the target subconcept belonging to one cell we need to specify an aggregated parameter or measure as well as the aggregation function. This parameter can be any item property including derived properties. After that for each cell we can find its aggregated parameter value.

If concepts `D1,D2,...,Dn` define dimensions and their common subconcept `T` defines the set of aggregated items with property `m` as a measure then the query `{d1:D1,…, dn:Dn}<sum({t:T | T.d1=d1 & ... & T.dn=dn}<t.m>)>` will return a result concept `n+1` dimensions where first `n` dimensions determine the cell coordinates in the input space and the last dimensions is equal to the aggregated property.

In order to change the level of details we need to simply move from the current concept to either some superconcept or some subconcept. If we choose a superconcept then this means that we will have less items and less cells in the final representation. Such a representation is more general since it hides some details. If we choose a subconcept instead of the current dimensions then it has to lead to the target concept and the number of details will increase.

If one of the dimensions reaches the target concept, which has the aggregated measure then this will mean the maximum level of details. Such a representation hardly makes sense since obviously no aggregation will be performed because each cell will contain one or no target objects.

Frequently we want to consider only a subset of all items in the dimensions, i.e., only a part of the whole OLAP cube is interesting for analysis. For example, we might choose an interval in time or only one product. Notice that the restrictions need not be imposed directly on the dimensions chosen for analysis. These could be other dimensions in the schema or sub/superconcepts. In this case we simply add these restrictions as the query constraints.

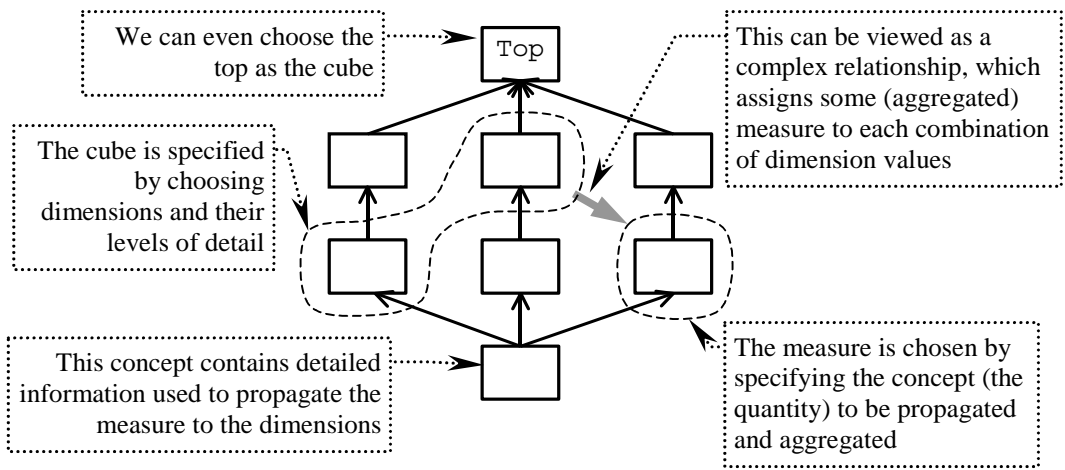

*Figure 16. To build an analytical representation it is necessary to choose dimensions each with some level of details (which can be varied by means of roll up and drill down operations) as well as some measure. Using the lowest level items the measure values are computed for each combination of dimensions values.*

Below we summarize some distinguishing features and emphasize advantages of the concept-oriented approach to analytical processing:

- There is no separation between the relational model for storing the original semantics and data cubes for defining structure and rules of aggregation. The concept-oriented model is designed in such a way that it is intrinsically hierarchical and multidimensional retaining all features needed for online transaction processing (OLTP).

- Dimensions and measures are symmetric and have equal rights in the representation. In particular, we aggregate dimensions along measures.

- Dimensions themselves can be multidimensional while measures can be multidimensional and hierarchical.

- Dimensions may have a common structure by using common concepts. In this sense it is not possible to separate them because their hierarchies intersect.

## 4.6 Inference

Concept-oriented database model consists of syntax (schema) and semantics. The syntax defines the state space, which is a set of all possible points where points are combinations of attribute values. The semantics selects a subset of all points defined by syntax, which are declared really existing (or true in logical terms). One specific feature of the concept-oriented model is that the syntax is hierarchical because attribute values can be new attributes with their own values, which in turn are normal object described by their own attributes. The syntactic points in this case may be defined on different levels, i.e., there exist large coarse-grained points and small fine-grained points. One approach to select conventional (flat) multidimensional spaces with different scales is defining a multidimensional cube (OLAP technique). The semantics is therefore also hierarchical and is distributed across levels of details. For example, we can consider combinations of countries and product classes at higher level or administrative districts and concrete products at lower levels of abstraction and in each case we can assign some semantic characteristic to each combination of values.

The idea of inference in such a formal setting is that we can impose constraints on some attributes and then using the existing semantic constraints stored in the database find the resulting constraints on some target attribute(s). For example, as we restrict a set of possible countries, the set of products produced is being also restricted (such an inference is known to be monotonic). This approach is rather general and concrete logical inference mechanisms depend on data and knowledge representation models, types of constraints and other details. For example, we might choose finite valued attributes and two-valued or fuzzy semantics. If in this case there are two attributes `Size` and `Colour` taking values {`small`, `large`} and {`green`, `red`}, respectively, then the state space consists of four points {`<small, green>`, `<small, red>`, `<large, green>`, `<large, red>`}. However, the knowledge may restrict the number of possible combinations of values. If so then restricting values of one variable will result in restricting values of the output variable because some their combinations are known to be impossible.

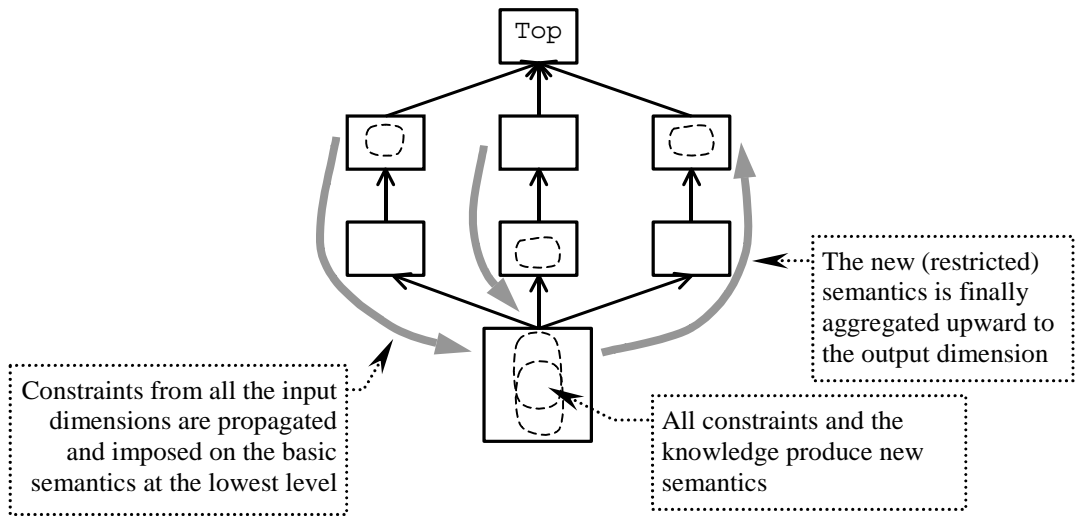

*Figure 17. Inference in concept-oriented database model. Constraints from all input dimensions are propagated downward where they are combined with the knowledge and the result is then propagated upward to the output dimensions.*

There are two general classes of knowledge representation: extensional and intensional. The first approach represents constraints on possible points explicitly, for instance by enumerating all possible points (necessity semantics) or all impossible points (possibilistic semantics). The second approach represents large intervals of impossible points while all other points are supposed to be possible. Suppose that we have never seen small green objects so they are considered impossible. This knowledge can be represented in the database by inserting tree possible records. After that we can carry out logical inference by imposing different constraints on one attribute and deriving new constraints on the other attribute. Normally the input constraints describe some concrete situation or object with certain properties. In the above example if our object is known to be small then we can easily derive that it cannot be green. Otherwise this object would contradict to the knowledge stored in the database.

The inference mechanism of concept-oriented database model generalizes this approach onto the case of hierarchical attributes used by database syntax and knowledge representation by means of concept items. The idea of such an inference consists in (i) propagating input constraints from each attribute downward to the lowest level, (ii) imposing these constraints onto the semantics on the corresponding concepts, and (iii) propagating this new constrained semantics upward to all target attributes (Figure 17).

Suppose that we want to choose only a subset of all product categories as well as an interval in time the products were sold. The task is to derive a set of countries in which the products from these categories were sold during the specified time. For that purpose we impose constraints on two input attributes, then propagate these constraints downward to the concept that combines products, time and countries, and finally find the list of countries by aggregating information upward.

## 4.7 Multidimensional Hierarchy

The multidimensional hierarchy called also analytical representation is a method of data modeling and viewing, which combines hierarchical and multidimensional types of representations. We can view the data as a tree, as a table or both. In addition we can configure the representation so that the tree and table structure may change.

For example, suppose that we have a hierarchical list of tasks to do. In addition to be characterised by its parent each task has a number of additional properties such as project, manager, some dates etc. However, on the other hand it is important that these properties themselves may be hierarchical, e.g., people assigned to the task in one or another role are hierarchically ordered according to their position, e.g., manager may have his own manager. Projects constitute their own hierarchy. Dates have a built-in multidimensional hierarchical structure and so on. One way for representing and manipulating data might consist in choosing tasks as a primary hierarchy while other elements are considered as dimensions. In this case we insert new tasks as children of some other tasks with simultaneous assignment of other properties. However for analysis it may be more convenient to view a hierarchy of projects or dates where the task is one of the properties. Or we can switch to a flat table view with a set of selected properties. Or we might want to view an administrative hierarchy where projects, tasks and other elements are characteristics.

In more complex cases the hierarchy may include internal tables as its nodes, i.e., the child items can be characterised by a set of properties and may have their own children with their own attributes. This view combines features of both trees and tables and could be called a table tree (tree of tables) or tree table (table of trees). One simple case can be obtained from a tree with nodes represented by tables. Another (dual) case is where rows of a table can be expanded into a set of children.

It should be noticed that the type of representation chosen does not change the semantics of data and its schema in the database. Different representations are based on formally defined transformations between multidimensional and hierarchical views. For example, if we build a view with tasks as a tree and people as its characteristics then it does not mean that in the database schema we directly assign a task to each person. The problem is that we might well view the semantics as a tree of persons with tasks as characteristics. So there always has to be some representation that is independent of the view. One approach to building various views from one and the same concept-oriented database schema is based on describing multidimensional hierarchies.

The multidimensional hierarchy is based on the principle "reference as inclusion", i.e., each item belongs simultaneously to several categories, which are its dimension values stored by reference. Thus each time we assign a value to an item attribute we put it into the corresponding category represented by the value item. (Such an interpretation of references is a specific feature of the concept-oriented approach, i.e., they cannot be used freely with no constraints like in programming but rather they have a concrete meaning for the database semantics.) Later we can consider this value as a parent node and all items that store it as a value will be considered its child nodes. The categories each item belongs to may themselves have rather complex structure. In general case we cannot select in advance the role of each dimension, i.e., if a dimension is a primary category or a secondary one or it is a characteristic of other items shown in the tree. All concepts in the database schema are symmetric with no role assigned in advance. The role is assigned only when we need to produce some view and this role is made explicit in a query.

A multidimensional hierarchy is a tree where nodes are items, which can be alternatively expanded into sets of other items. Here are the main features of such a structure:

- The nodes are multidimensional items rather than simple entities, i.e., they are characterized by several attributes and can be represented as a table rather than a list.

- The characteristics of nodes may be complex properties retrieved via queries from other parts of the database. In particular, they can be multi-valued properties.

- There exist alternative expansions, which represent alternative views on the item composition, e.g., each project can be viewed as consisting of personnel involved into it or a list of allocated resources or departments involved into their execution.

- A possibility to define the expansion by means of complex queries or multi-valued properties.

The root of any analytical representation always corresponds to the top concept, which represents the whole problem domain at the lowest level of detail and highest level of abstraction. For such a representation there are no details or characteristics at all. For each node type in the analytical representation it is necessary to specify a set of alternative expansions, which correspond to different views on this node type composition. The first level of the analytical representation is defined by specifying the concepts that will be shown when the root node is expanded. For example, we might specify that at this level we view our problem domain as a list of projects, persons or resources. Then for each of these concepts we have to specify their own alternative views, which represent their own composition. For example, for projects we ask what is the possible composition of each project? If we want to represent each project as a set of people involved into it then we have to define the corresponding composition as a relationship or multi-valued property via query. Generally in order to define a multidimensional hierarchy we have to specify a path from the top concept to some lower level concept (Figure 18). For each concept in the path we have to define a set of properties to be shown and a set of alternative expansions, which a possible path continuations.

This technology provides very flexible means for data representation. For example, we can use hierarchical categories such as internal projects or product categories in order to select data in other concept. In this case we might select one intermediate category such as a type of product and then expand directly into some target concept items, e.g., personnel involved into its production. Or vice versa we might select a person and then expand it into a list of project product categories or high level projects it is involved. Interestingly we can continue this expansion process and even view the same concept at different levels of the structured hierarchy. For example, after selecting people involved into a project we might again expand a person into a list of projects where he acted as a manager. Notice that here we used the same target concept (projects) but with different relationship.

This multidimensional hierarchy representation method is especially useful when augmented with item filtering mechanism. We can simply restrict the number of items shown on each level so that it fits our needs. However we can impose much more complex constraints on arbitrary dimensions and they will be automatically propagated through the concept-oriented structure. For example, we can select only some departments and then the view will automatically exclude all other items (projects, people, resources etc.), which are not associated with these departments.

The analytical representation is also very helpful in navigating the data. Search for information is a very actual problem. The traditional approach consists in using the item contents to find necessary data. An alternative approach consists in using relationships among items to search for information. Using the analytical representation we start from very general representation and then narrow it down by choosing appropriate expansion directions and imposing filters. Thus the target items are found by following the appropriate relationships.

Instead of the interactive navigation through the analytical space we can use an automated search. For that purpose we have to formulate the relationships the searched item has to satisfy. For example, an item has to belong (directly or indirectly) to some type of project, it has to be created in the second quarter of this year and is somehow related (we do not how) to the specified person.

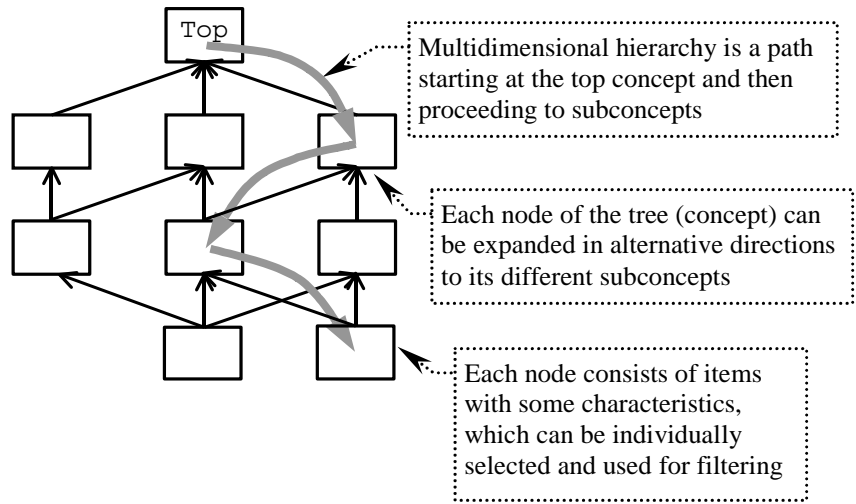

*Figure 18. Multidimensional hierarchy is a path starting at the top concept and then proceeding to its subconcepts. The tree shows items of each concept in the path with all selected properties, which can be used for searching, ordering, navigation and filtering.*

The method of describing is very similar to top-down development of the database schema where we start from general concepts and then define their subconcepts. In multidimensional hierarchies the idea is that each concept may have alternative views, which describe possible item compositions. For example, a country can be considered as consisting of administrative districts, companies or people; an organisation can be considered as consisting of departments, projects or personnel. Such alternative views are introduced as subconcepts for the root concept.

Sometimes new subconcepts representing alternative views involve many already existing superconcepts. For example, we might think of personnel of an organization as consisting of projects because each person can be expanded into a set of projects assigned to it. Yet the concept of projects already exists because the organization consists of a set of projects it executes. However we can notice that projects can be also viewed as consisting people assigned to them. The solution is that we introduce new subconcept as new alternative view of personnel however this concept simultaneously is an alternative view for the project concept. This subconcept inherits properties from both superconcepts and thus can be serve two views depending on what is expanded.

By adding new subconcepts representing alternative views of existing concepts we introduce new levels of detail because each such subconcept contains elements of already existing items. Without these new subitems the description is more general because we cannot say what each item consists of. Thus the more subconcepts we have the more detailed description we can get by expanding items and getting their subitems. It is important that concepts can be expanded in many alternative directions. The bottom concept contains most of the details existing in the model while the top concept represents the model as only a single item. For example, the top concept is an organisation as one element without internal elements or one country without internal contents.

## 4.8 Self-References

As defined in the concept-oriented database model the loops and cycles in the schema and item reference structure are not allowed. In particular, it is not allowed for a concept to have a dimension with the domain in this very concept. However, we in practice there is a lot of cases where such self-references look quite natural and at least can be used to compactly represent relationships among items. For example, suppose that the personnel are stored in one concept. Each person has a manager, which is an item from this very concept. Thus we need to create a dimension with the domain from this very concept.

In this example it is important to understand that the ability of items to be characterized by items of the same type is a separate relation or characteristic, which formally must exist as a separate subconcept (the principle of relations as lower level objects). This separate subconcept is more specific because it combines two elements: the source item as a person and the target item as a manager for the person. Such an explicit representation is formally correct but practically inefficient.

The main rule is that if we want to characterize items by more general items (from a superconcept) then we introduce a new dimension. If we need to characterize items by items from any other concept (not a superconcept) then we introduce new subconcept to implement this relationship. In particular, if we want to characterize items by these very items then we formally have to introduce a new subconcept for that, which represents the corresponding relationship.

Thus when new characteristic is not taken from a superconcept we must introduce a new subconcept to implement this type of characterization (Figure 19). However, in the case of many-to-one relationship where the characterized item is always one item from the domain we can easily optimise the representation. In this case for each characterized item there is precisely one item in the corresponding subconcept, which is an instance of the relationship. This item in the subconcept stores the reference to the characteristic. Thus we can simply combine these items and store them in one original concept. For self-characterization this means that we simply store a reference on this very type of concept. In general case we store a reference to one item from an arbitrary concept in the schema. For example, if one email belongs to one project then we can store the project reference directly in the email item although it is formally not allowed because projects are not stored in a superconcept of emails.

This optimisation can be applied to any many-to-one relationship where an item is characterized by only one item from any other concept. The database management system should distinguish such optimised representation from the true characterization of items by their superitems.

One consequence of such a simplified representation of relationships is difficulty in future schema alterations. The problem is that such changes in the database schema may easily lead to the necessity of having an explicit relationship as one separate concept. In this case we need to remove the corresponding dimension with self-reference and create the corresponding subconcept.

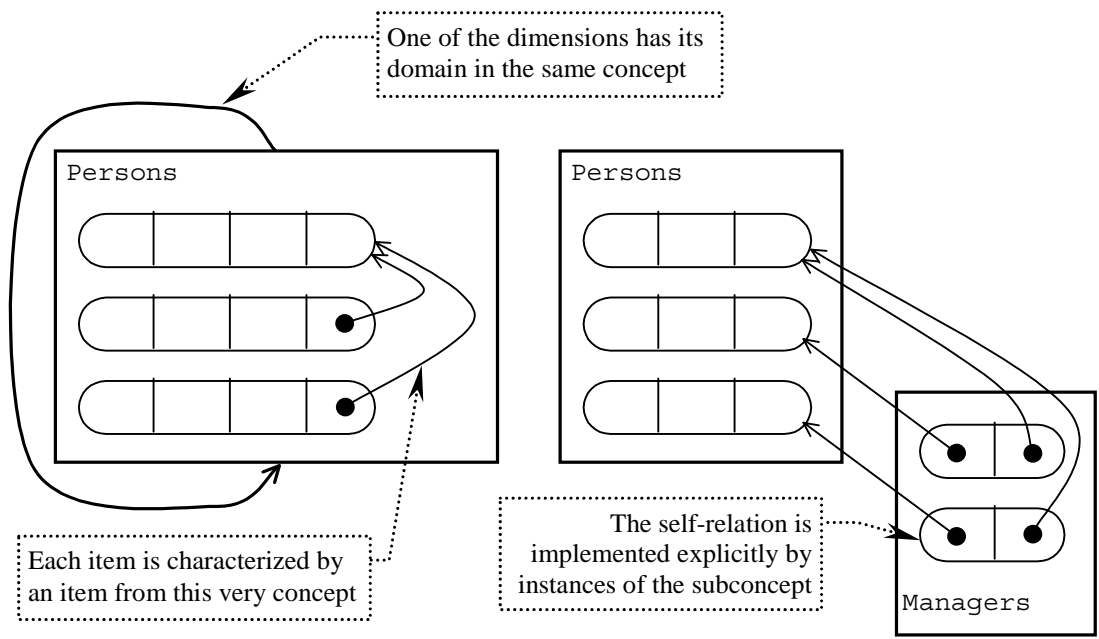

*Figure 19. Formally self-references have to be implemented via a separate subconcept, which stores instances of the relation. However for performance reasons any many-to-one relationships can be implemented directly by storing target reference in the source item.*

In some cases two concepts are symmetrical in their mutual role. Frequently they simply represent parts of one object so that they have the same life cycle. For example, concepts User and ContactInfo might have one-to-one relationship so that for each user there is precisely one contact information item and vice versa. Theoretically we could say that contact information is more general concept while user is characterized by some contact information. Indeed the contact info can be hardly characterized by user. In this case we might specify user as a subconcept of contact info so that each user item stores a reference to the corresponding contact info item. If users may have several contacts then we need to introduce a common subconcept for them and then we clearly see that these two concepts have the same level. And this is why we might simply store a reference from contact info item to the corresponding user item.

Such situations appear when it is difficult to establish a concept-subconcept relationship. For example, for concept Groups and Users it is clear that each group consists of a set of users. However, in other situations the relationship may depend on subtle implicit assumptions about the nature of the problem domain to be represented.

## 4.9 Graphs and Trees

Frequently the problem domain is naturally described by a structure, which is similar to graph or tree. This means that the model consists of nodes (of one or more types) and edges (also of one or more types). A direct approach to such a problem could be to associate nodes with concepts and the concept instances with other nodes, which are somehow connected with the parent node. Here the idea is to exploit the "instance of" relation between instances and concepts to represent some relationships between nodes of the complex graph. Such an approach looks particularly attractive for representing trees.

However, it is not the best ides and this method hardly can be called appropriate for representing graphs (and even for any other problem domain). For the general point of view we basically do not want to exploit the "instance of" relation between items and their concepts for representation purposes. It could be considered an internal mechanism used for the database model itself. When modelling some problem domain we strictly separate syntax and semantics including separation between their elements and relationships. In other words, we have concepts with their own subconcept-superconcept relation and we have items with their subitem-superitem relations. These are the only facilities that can be used to describe a problem domain. The syntax is used to represent the problem domain space structure, i.e., the world itself with all possible states. The semantics is used to fill this complex space with some objects by assigning them concrete coordinates. Notice that the syntax is the stable part of the description, which is not supposed to change frequently (just like any coordinate system) while the semantics is floating or flexible and is supposed to be regularly updated to reflect the latest state of the problem domain. In this sense their roles are completely different and before we start modelling some problem domain we have to explicitly specify what kind of elements will be represented as syntax and what kind of entities will be treated as semantics. Sometimes the choice is straightforward but in many cases it is not so obvious.

When modelling graphs and trees it is important that both the nodes and edges belong to the problem domain semantics. Indeed we should be able easily add and remove both nodes and connections among them. In this sense nodes and edges behave like normal objects. Particularly, they may have different types and properties. If we associated them with syntax then we would have to frequently add and remove concepts by changing the space structure. However concepts are not objects and we cannot manipulate them like objects. In particular, they do not have references or properties. The only reason why we would like to represent nodes by concepts is that the concept structure will look precisely like the modelled tree or graph. Yet this is a wrong way.

To represent any graph we start from describing its nodes, which have a primary importance because they exist before relations among them. Essentially we apply here the principle of relations as objects at lower levels. According to this principle nodes exist before their relations (edges) and have a higher level with respect to them. In the case of one node type we simply introduce the corresponding concept the items of which will be concrete nodes. As the semantics changes the nodes are added, removed or updated. Notice that according to another concept-oriented principle nodes do not know about relations among them because relations are objects at lower level, which are not visible. The nodes then cannot store any references to edge objects. It is quite natural (and very important) because we do not know in advance how many and what kind of relations will exist and will be added after the nodes are defined.

Once nodes have been described via their own concept we can introduce relations among them, which are edges of the graph. If edges have one type then we simply define a new concept for them where one edge is represented by one item. Notice that this concept has to be a subconcept for the nodes. The edge concept must have at least two dimensions both with the domain in the node concept. The first dimension indicates the source of the edge and the second dimension indicates the target. Of course there may be more properties if needed. If there exist more than one type of nodes or edges then this problem is being solved by type modelling. For that purpose we should use more complex characterization schema. This means that nodes and edges are positioned in a more complex space with their own hierarchies and dimension structure.

In general case we can represent even more complex structures than simply graphs and trees. For example, we might use hyperedges, which connect more than two nodes. However, such a case is difficult to represent via one subconcept with fixed dimensionality because the number of connected nodes is variable is not known in advance. For that reason it would be not very natural to encode the maximal edge rank into the syntax. Indeed the number of nodes is not limited while the concept dimensionality is fixed. To solve this problem we can represent each edge as one item with general common properties while the nodes it connects are stored as separate items in another subconcept. Thus the edge item will not know the list of nodes it connects because they are stored in a subconcept. Of course we could always define for convenience some virtual properties such as getting for an edge its list of nodes.

Another more complex example is where edges themselves might have some relations or associations. In simple cases it could be implemented via more complex characterization. For example, each edge could be assigned a category as some superitem. In more complex cases the relation could be implemented via subconcepts.

## 4.10 Inheritance Modelling

Although inheritance is not explicitly used in the concept-oriented model there are some similarities in building a representation via inheritance relations and the concept-oriented approach. The main idea of inheritance is that the extension gets all properties of the base object and such an approach is used mainly to define more specific elements from more general ones and this is why only single-inheritance is normally used in practice. In this sense it is obvious that subconcept are analogous to extended more specific classes while superconcepts correspond to more general base classes. However the way how this relationship is used is significantly different.

In the object-oriented model instances of extensions store all fields of all the bases. In particular, creating an instance means creating all the bases, which are embedded into this extension. It is important that only the extension has an identity in the form of a reference while all its bases are included by value where they loose their individuality because they are treated as fields of the extension. In the concept-oriented model extensions (subitems) store only their own fields, which are essentially references to all the bases (superitems). It is important that all the bases (superitems) are separate objects with their own identity and life-cycle. In particular, the base items can be shared among more than one extension of one or different types and we can manipulate them separately (Figure 20).

Generally the interpretation of the concept-oriented relation between concepts is different from that between classes in the object-oriented model. Concepts are created as independent entities, which are described by introducing dimensions. Each dimension is used to characterize instances of this concept by instances of the domain concepts. Obviously here we do not have anything from inheritance. In particular, the number of dimensions is normally rather high and all the dimensions have equal rights so we cannot select some domain as a base concept in object-oriented terms. For example, it would be in the style of object-oriented approach to extend the class `Wines` by the class `RedWines`. The red wines could have additional fields, which are specific to only this class of objects, which are added to the inherited fields of the base class of wines. Then each instance of red wine would be one object with all the inherited fields. On the other hand we would have no separate objects so the class of wines would be empty.

Such a style of model specification is not typical for the concept-oriented approach. The difference is that when we introduce a new more specific concept such as `RedWines` we are not thinking about reuse or inheritance. Instead we are trying to understand what are the characteristics of this new class of objects. We might well choose the concept `Wines` as one of characteristics but it does not mean that we want to inherit its fields in the object-oriented manner. We simply want to assign a single superitem as a parent for each new subitem. These parent items then can be interpreted as a group/collection/set this item belongs to or its coordinates so we bring a completely different semantics into this relationship. Interestingly, if we have only one superconcept then the subitems are essentially empty because they store only one reference to their superitem with no other characteristics. This is why in the concept-oriented approach there is normally many superconcepts.

The concept-oriented approach introduces new interpretation of the subitem-superitem relation, which is absent in the object-oriented approach. The object-oriented interpretation can be reduced to reuse of the base object. In the concept-oriented method it is much more rich and expressive because it is interpreted as inclusion relation, which is the base element for the whole model.

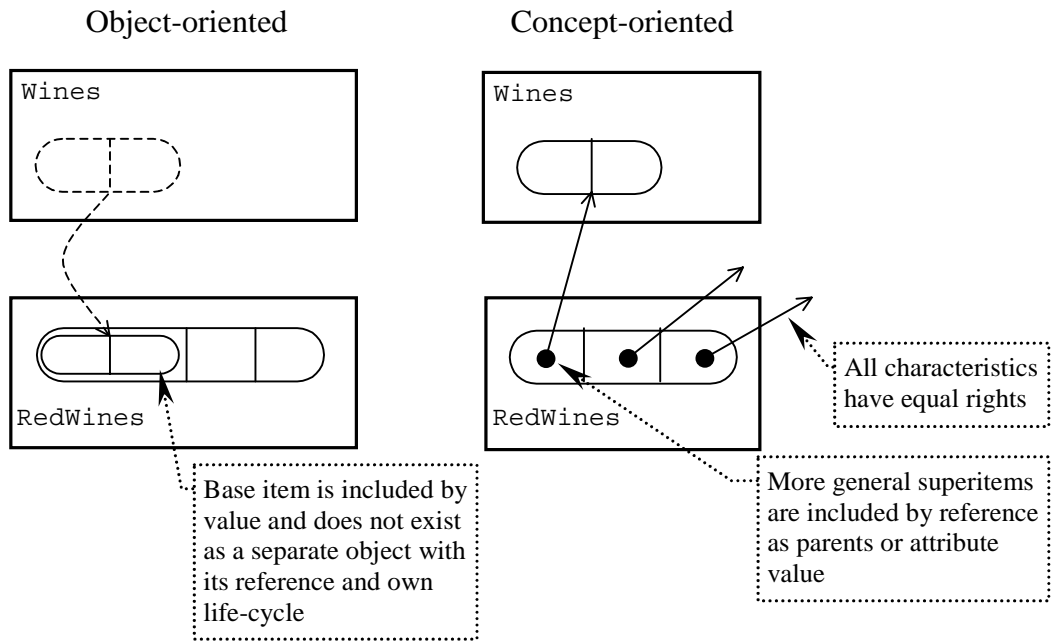

*Figure 20. In the object-oriented model base items are included by value into the extension and loose identity while in the concept-oriented model the more specific superitems are included by reference as individual objects.*

## 4.11 Type Modelling

This problem originates in the object-oriented paradigm where elements are assigned classes, which provide a structure for its instances. The classes then can be extended and the new more specific classes inherit all properties of the base classes. The idea is that when an object is being created we specify its class and then its structure is created automatically. It is important that all classes form an inheritance hierarchy, which allows reusing their structure and functionality.

Such style of thinking is not typical for the concept-oriented approach where classes do not exist in the form they are used in the object-oriented paradigm. The classes have at two functions: instance managers (creating, containing, life-cycle managing etc.) and structure managers. The first function is typical for object containers. The idea here is that we define a class as unique container and then all objects within this class obey certain discipline. In particular, all instances are provided a common structure and functionality. This function is very similar to that of concepts in the concept-oriented approach and the connection between them is described in section 4.10.

The second function is used to describe types of containers themselves. For example, we might describe a type `Address` and then define several containers (concepts, tables etc.) of this type such as `PersonAddress` or `CompanyAddress`. Notice that the type here is not a run-time object but rather is interpreted as a pattern used to create run-time real elements like containers and their instances. Such an approach is used in object-relational databases where tables can be assigned a type.

If we are using separate types as patterns for creating containers (concepts, tables etc.) then it is simply a convenience tool. Indeed, we can easily describe a concept structure on some language, then provide a nice name for it and finally use it as a shortcut when creating new concepts. The problem is that types can be extended. For example, suppose that we have a generic type of persons and then we find that we need to describe a more specific type of persons with unusual characteristics, say, managers. Managers are obviously normal people but on the other hand they have additional characteristics. One approach is to create a new type by extending the base type of persons.

Once such an extended type has been described it can be used to create real containers. In particular, we might create a concrete table or concept for storing information about managers. Instances of such a concept would inherit all properties of normal people but kept in a separate container. Obviously here we again obtain the situation where the base object looses its identity and is included by value into its extension. Moreover such a container created by using an extended type will have no connection to containers created using the base type. Indeed keeping such an association even formally is hardly possible and does not make sense because there may be several base type containers and several extended type containers. Another problem is that for each new extended type container we have to specify its concrete dimension domains but it is not possible to do because there may be several instances of container of the base type.

So the inheritance relation stays outside the run-time container structure and should be considered a convenience tool for describing container structure. In other words, the type inheritance hierarchy is a separate structure, which makes it easier description of new containers by reusing already existing descriptions. This inheritance does not have any influence on run-time structure or functionality of containers once they have been created. For example, we might have a type for a container, which stores media library entries. Each user can then create its own media library by using this type. The system then will manage many instances of containers of one type. However the type hierarchy has no relation to the concrete run-time concept structure of the database.

## 4.12 File System

A very interesting topic for any database model is its relation to the file system. Indeed both file systems and database systems are widely used to store information but on the other hand they are still separate technologies based on different paradigms. Advantages of file systems is simplicity and the presence of hierarchies while advantages of database systems are in their ability to relate fine grained pieces of information. In this context the challenge is to create a paradigm and technology, which could combine these two approaches to information storage and management. Surprisingly such an extremely actual conception does not still exist. In such a combined system we would like to manipulate files like database items by relating them with each other and other items. On the other hand we would like to keep concepts in a hierarchical folder structure and treat them like elements of the file system.

Such a combined technology can be created on the basis of the concept-oriented database model. In this model we first of all suppose that files are items of the special (system level) concept called `Files`. Thus each file gets its reference and a set of characteristics. The reference is essentially the file handle, which can be stored in other items as the file representative. The file characteristics include the conventional attributes like the date of creation and modification, size, permissions etc. One of the file item characteristics is its user defined contents. This characteristics is normally has a special implementation because the files may have large size and need special interfaces to access their contents. Yet from the database organisation point of view it is just another file attribute. For small files this characteristics should be stored by value while for larger files it may store some lower level (physical) identifier of the corresponding object (like a file handle in the underlying file system or a starting block number).

Such an integration of files into the database model has an immediate advantage. Now we can create dimensions of our concepts having the domain in the `Files` concept. One approach consists in extending interpretation of one or more of the files by adding our own information. The file in this case should not be processed directly because it has a higher level information attached in the corresponding superitem. For example, an application might maintain a list of emails stored in our own concept but long attachments could be stored in separate files. These files are declared as dimensions of our concept and then accessed via the database system like any other item. Or an application might manage multimedia information and in this case again it might store long media files as items in the `Files` concept and then store their reference in its own concepts along with other user information and associations.

Folders of the file system can be also represented as items of a special (system level) concept called `Folders`. Each folder has a set of properties similar to those of files and just like any other item in the database a folder a reference, which can be used to represent it in other items.

Once we have folder items we can use them to include into them any other sort of item. In particular, we can include files into folders. The simplest way to implement this relationship consists in using a special dimension for files where a reference to the parent folder is stored. If files are allowed to have more than one parent folder then we will need a special lower level concept for that. Each folder is also included into some other folder. This is obviously a self-relation, which can be implemented by allocating a dimension in the `Folders` concept or by creating a special concept. In any case file and folder items know their parent folders (while folders do not know directly their members).

By introducing `Files` and `Folders` concepts we provide legal status for file system elements, which now can be used like any other database item. In other words, we can create and manipulate file system elements like items from our own concepts. (There should be some constraints because these items have a status of system objects, e.g., these objects might require higher permission for special operations.) The next task consists in make it possible for treating database elements like file system objects. Practically this means that if we browse the file system that we have to be able to access information in the database elements. For that purpose we should include each concept into some folder. After that each folder will consist of three types of elements: internal folders, files and concepts.

In order to implement such a relationship we can introduce another system concept called `Concepts` (Figure 21). Each item of this concept represents one user concept with all necessary attributes including the parent folder. Here we clearly see the necessity of the system status and special treatment for these three special concepts (`Files`, `Folders` and `Concepts`). They are analogous to system tables and catalogues of the relational database management systems. Indeed the concept `Concepts` contains items, which are themselves concepts on the user level of the system. It should be noticed that such an organisation is not a defect, technological trade off or a performance optimisation at all. It is the essence of the concept-oriented paradigm where all elements and functionality are spread across all layers of the system organisation and their role may well change when considering them at different levels. We described already such a phenomenon when postulated the principle of relations as object at lower level where the consequence is that an object may play the dual role depending on the level it is viewed from. Here we again see an example of such multilevel organisation by introducing new level system level of the database.

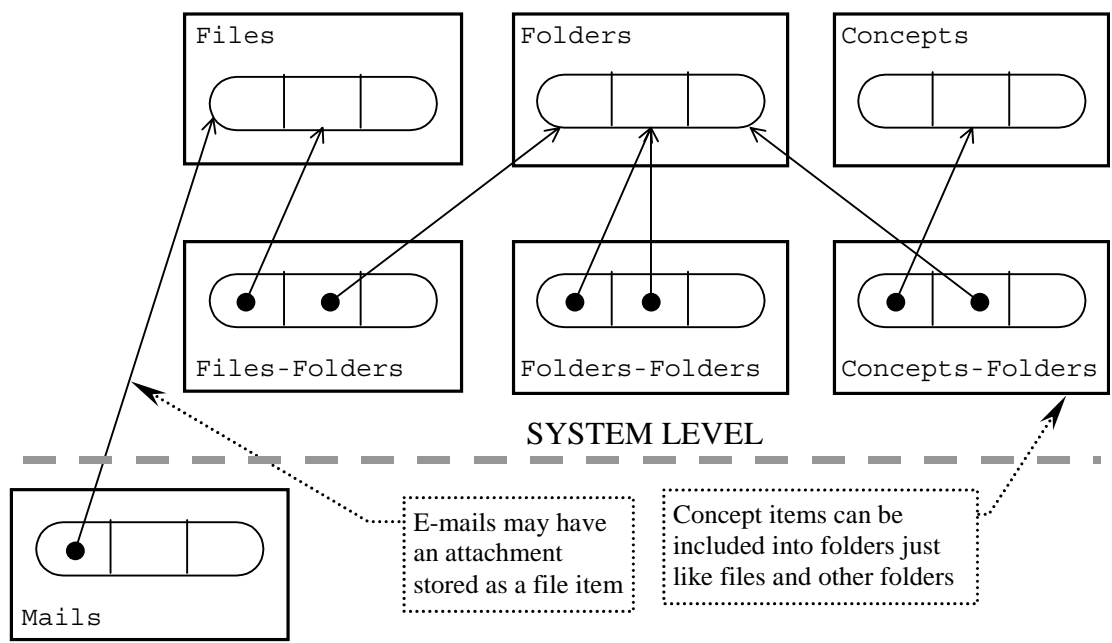

*Figure 21. Files, folders and concepts are items at the system level, which however can be used as dimension values in user concepts.*

It should be noticed that there is an alternative design for the integration of file and database systems. The idea is to represent folders by concepts where items are members of the folders. In this case we would have the folder structure represented directly by the concept structure. Such an approach again tries to exploit the syntax structure to represent the semantics what has been demonstrated not very appropriate. In other words, here we model the inclusion into folder relation via instance of relation between items and their concept. For example, as an item's parent we will get some concept rather than another item while we want to deal only with items as objects and properties. Although we find this approach unsuitable it is important to emphasize its existence as a design alternative (what we have also done for graph representation in section 4.9).

Notice that in the database system there may be many concepts of one type. For example, each user could create its own concept called `Contacts` all of which have the same syntactic structure. Such concepts are stored in different folders and may have different names. They also use different domains as the characteristics (although the domains have also the same type).

The name system plays significant role in any file system while in database system it is not so important because most of objects are identified by references or by other methods. In other words, in a file system names are the only way to access data elements because other identifiers may change in time. In databases only syntactic elements such as tables are identified by names while records are represented by primary key or similar identifiers.

It is however important to understand that the name system is a separate layer intended to identify objects of different types. It is a higher level identification system, which is based reference management system. Its target audience is human users. Its main function is quite simple and consists in assigning some internal (lower level) objects an external (higher level) name. We suppose that any item at system level may have some name assigned and there the corresponding level at which the items can be accessed via their names. This level's functionality is essentially reduced to creating names and their resolution.

In addition to serving the system level objects like files, folders and concepts the name system could be used also naming user items like individual records. On the hand it should not be obligatory to assign a name to all the available file system objects. This means there might well be files, folders or concepts without any name. Indeed most of the files in the contemporary software systems are used by some applications rather than by users. Why then these files need so computationally expensive means of identification as human readable names? For example, when we create a concept with user contacts we might want to provide some name for it in order for the user to load it in different applications or to copy to different computers. However, this concept may have as its dimensions files and folders where some additional information is stored. They are considered a part of the whole representation and these numerous small objects definitely do not deserve their own name because they are manipulated by applications only. Thus when creating files, folders or concepts we can choose creating a name as an option. Or alternatively names will be created only if objects are accessed via the name system layer of the whole system. If objects are accessed via the reference layer then names are not used.

# 5  Related Work

## 5.1  Database Models

### 5.1.1  Hierarchical Database Model

The hierarchical data model organizes data in a tree structure with a hierarchy of parent and child data segments. For example, an organization might store information about departments such as name, head of departments etc. The organization might also store information on workers in each department such as name, data of birth, salary. The department and person data forms a hierarchy, where the department data represents the parent segment and the person data represents the child segment. If a department has five people, then there would be five child segments associated with one department segment. In a hierarchical database the parent-child relationship is one-to-many. This restricts a child segment to having only one parent segment.

The concept-oriented database model naturally supports hierarchies and simultaneously overcomes the main difficulty, namely that there may be many different hierarchies produced from the same data. For example, if we consider a relationship between projects and persons then obviously we can build two hierarchies. The first one starts from projects and then for each of them enumerates persons involved in the projects. The second one starts from persons and then for each of them finds a list of their projects. In the hierarchical data model it is difficult to model such cases while in the concept-oriented model they are represented in a very natural and flexible way.

### 5.1.2 Network Database Model

The popularity of the network data model coincided with the popularity of the hierarchical data model both of which preceded the relational model. The network model was designed to solve serious problems of the hierarchical model. One of them is that many practical cases had to be modeled with more than one parent per child. The network model permitted the modeling of many-to-many relationships in data. The basic data modeling construct in the network model is the set construct. A set consists of an owner record type, a set name, and a member record type. A member record type can have that role in more than one set so it is possible to have multiple parents. An owner record type can also be a member or owner in another set. The data model is a simple network, and link and intersection record types (called junction records by IDMS) may exist, as well as sets between them. Thus, the complete network of relationships is represented by several pairwise sets; in each set some (one) record type is owner (at the tail of the network arrow) and one or more record types are members (at the head of the relationship arrow). Usually, a set defines a one-to-many relationship, although one-to-one is permitted.

In some sense the network model is very similar to the concept-oriented model because its high level view is also a multidimensional and hierarchical where children have multiple parents and parents have multiple children. However, the mathematical formalism behind this model is too complicated and this is why it was very difficult to implement, maintain and use. In contrast the concept-oriented model is designed to be simple with clear definitions of what is syntax and semantics.

### 5.1.3 Relation Database Model

In the relational database [1] the data and relations between them are organized in tables. A table is a collection of records and each record in a table contains the same fields. Certain fields may be designated as keys. Where fields in two different tables take values from the same set, a join operation can be performed to select related records in the two tables by matching values in those fields. It is important that the way how the tables of data are stored makes no difference. Each table is identified by a unique name and that name can be used by the database to find the table behind the scenes. Thus the only thing the user has to know is table name and some information about records to be retrieved such as primary key, column values or other criteria. It is different from the hierarchical and network models where the user has to know the data organizational structure in order to retrieve it, e.g., by following some path in the hierarchy. The main advantage of the relational model is its simplicity. Indeed in this model there only individual tables, which store records with certain structure and that is all. Given the powerful structured query language (SQL) it is possible to model almost any practical case.

The relational model is general enough to express almost any other data model. Yet the main problem here is that this expressiveness and flexibility is reached by means of lower level manual control of operations. For example, in the relational model there is no such a notion as inclusion of records although physically and logically it exists. For this reason the database does not manage correct relationship among these elements and it is the task of the user to maintain the model in the consistent state. In particular, the relational database does not know what is a lower level record or higher level record because all tables have the same status. However this hierarchy *objectively* exists in the problem domain and must be reflected in the database model in order to function properly. In this sense the relational model is very good tool for implementing physical storage for objects and very flexible, reliable tool with simple and clear conception for manual low level control and non-trivial manipulations. In particular, it is a good idea to use a relational database as a physical storage level for a higher level database based on the concept-oriented model.

Although the relational model has foreign keys it does not impose any structure on them and does not restrict their use. Just like in the object-oriented programming there are not constraints on reference structure and reference use, the relational model allows defining any structure for foreign keys. This is because foreign keys is a lower level mechanism with no higher level semantics. In contrast in the concept-oriented model the reference structure is the primary database construct.

The major difference between relational model and concept-oriented model is that the first one manipulates records in tables leaving the task of maintaining relationships to the user. The concept-oriented model on the other hand is responsible for maintaining and manipulating relationships among items, i.e., the structure of the database. For example, in the relational model it is possible to remove a record leaving other records, which have to be also removed in their tables. Or we may well have cycles in references. In the concept-oriented model it is not possible because it does not make sense according to the concept-oriented principles of the model. All subitems are automatically removed as soon as their superitem is removed and there is no way to bypass this procedure because managing relationships is the central function of such a database. In this sense the concept-oriented model appears more restrictive because it makes some assumptions about the structure of the world (of any problem domain) and this structure is then reflected in the database. The relational model does not impose such constraints and hence allows for much more freedom in interpreting the data and its semantics.

Below we summarize the main differences between the relational and the concept-oriented models:

- Relational database manipulates records while concept-oriented database maintains relationships and structure. Indeed in the concept-oriented model there is no such a notion as an isolated item because each item is a combination of other items so the database manages these combinations rather than records. Although the relational model has some mechanisms like foreign keys and integrity constraints, they are not finally conceptualized and can be used (or not used) arbitrarily as convenience techniques.

- All tables have equal rights, the same status and the same level so it is a flat model while concepts are organized into a lattice and the database knows about this lattice and one of its major functions is maintaining this lattice as the primary database syntax structure.

- Records in tables are connected implicitly and database does not know how they relate while in concept-oriented database items are explicitly linked by references and the database is responsible for the consistency of this semantic structure.

- Relational database does not have a canonical semantics. Particularly, we cannot semantically compare two databases. The relational semantics depends on the interpretation of table and their roles. For example, if there are two tables with the same column, say, `OrderNumber`, then the database does not know that these tables are somehow related. It is the user who writes queries knows how to interpret data in order to produce meaningful results.

- Relational queries are based on join operations with arbitrary join conditions specified in each query. So the database does not know about most of table connections. Concept-oriented queries are based on manipulating collections using the dependencies between concepts declared in the database schema. The concepts simply cannot exist without explicit relationships with other concepts, i.e., any concept is a part of the whole database schema.

- In the concept-oriented model the schema plays the key role in all operations, i.e., by definition any concept exists in connections with other tables. If we change these connections then the database structure and its semantics also change so it becomes another problem domain. In the relational model most operations are described manually.

- One specific feature of concept-oriented approach with respect to the relational data model is that multi-valued properties or attributes have a fundamental explanation proceeding from the inclusion relation among items. In relational databases such properties, which widely used in practice are either embedded as special type or have to be implemented manually via joins.

### 5.1.4 Object-Relational Database Model

Object/relational database management systems (ORDBMSs) add new object storage capabilities to the relational systems and is considered its evolutionary extension. The new facilities integrate management of traditional fielded data, complex objects such as time-series and geospatial data and diverse binary media such as audio, video, images, and applets. By encapsulating methods with data structures, an ORDBMS server can execute complex analytical and data manipulation operations to search and transform multimedia and other complex objects.

The object-relational model introduces some object-oriented features but is normally not qualified as a complete model. The thing is that this approach is considered a compromised between the relational model and the complete object-oriented approach. In this sense it can be considered a layer over the relational database. Such an approach is very practical since it allows using both the pure relational features and some object-oriented facilities. It should be noted that the object-relational model does not introduce its own semantics or paradigm. The most important enhancements are user-defined type and user-defined functions. Using these new mechanisms it is very convenient to model the problem domain by means of relational facilities and then access and manipulate data items as they were objects with their identity and methods.

### 5.1.5 Object-Oriented Database Model

The object-oriented database model is very progressive approach targeted at transferring object-oriented methods and approach into the world of databases. Unfortunately, in its current form this model took too much from object-oriented programming and could not select the most important from database perspective moments. Thus the object-oriented database can be considered an application with persistent state and because of this they lost performance, efficiency and what is more important database uniqueness and specificity. The object-oriented database does not distinguish between upward and downward references, i.e., parent/child relationships. The object-oriented model deals with and manages references while concept-oriented model deals with and manages inclusion relation among objects. In the object-oriented paradigm there are no constraints on object reference structure and the object-oriented database model objects in addition are persistent. However, this is no so in the concept-oriented model where objects (concept instances) are much more semantically rich constructs with the concrete problem domain interpretation of their reference structure. In this sense the object-oriented database model is an (naive) attempt to make conventional objects persistent while the concept-oriented model brings a new semantics and new representation technique in the world of data.

One of the main differences between object-oriented and concept-oriented approaches is the use of references. Object-oriented model allows for arbitrary references among objects while concept-oriented model allows an item to reference only items from superconcepts. Particularly, cycles are possible in object-oriented model and are prohibited in concept-oriented approach. This follows from different role of references. In object-oriented databases it is a universal mean for accessing objects while in concept-oriented database model they reflect the structure of the problem domain and particularly are interpreted as inclusion relation among items. A restricted use of references can be viewed as a serious constraint of the concept-oriented approach. However, it is not so. By changing the role of reference from the conventional access tool to structure representation we change the paradigm. Indeed, in object-oriented approach we can retrieve a reference then access the object and again get some reference so that there are no constraints on the path. In particular, two objects may store references on each other producing a cycle. So what is wrong in this approach? First of all in practice such a complete freedom entails numerous difficulties in reference and object management. However, the main difficulty is that access function of references is not the primary one in databases as conventionally assumed in object-oriented approach. Databases reflect the syntax and semantics of the problem domain and references play a new role where they are interpreted as relationships among items (inclusion, knows, attribute-value etc.). In other words, in concept-oriented approach we use references where we want to establish the fundamental inclusion relation between two items. And this is why a parent item may never store a reference to its children. It simply does not know if it may have children and what kind of children may appear in the future. Obviously this new role of references is fundamentally different from that used in object-oriented paradigm.

Another feature of object-oriented approach is the use of inheritance. In concept-oriented approach inheritance does not exist. At least it does not exist in its conventional form as a mechanism of extending class descriptions. Each new subconcept is described by specifying a set of its superconcepts so that each subitem will be equal to a combination of the corresponding superitems. Obviously in such a form it is not an inheritance (although we do not exclude other interpretations). Thus in object-oriented approach inheritance play a primary role as a mean of class construction while in concept-oriented paradigm such a role is played by the mechanism of specifying a set of superconcepts. Particularly, in most cases a new subclass has only one base superclass and only in some complex cases we need to more classes but normally only a few. In contrast, in concept-oriented approach a new subconcept normally has many superclasses and only in rare specific situations we might use one or two superconcepts. Notice that it is quite normal in object-oriented approach to have a class with no base class while in concept-oriented model a concept with no superconcepts is meaningless because its items will have no information (no fields).

### 5.1.6 Other Data Models

#### Deductive databases

Deductive databases are intended to represent knowledge and carry out logical inference. Their design strongly relies on logical programming and predicate calculus. Just like the object-oriented database model is an attempt to expand object-oriented paradigm on the area of databases, the deductive databases can be considered an attempt to expand logical programming on the area of databases.

The concept-oriented model provides a natural mechanism for logical inference. This mechanism is based on the notion of constraint propagation over the database concept structure. In contrast to deductive databases the inference mechanism is quite simple and is very well integrated into the whole model.

Associative Model

The associative model [2] divides the real-world things about which data is to be recorded into two sorts. *Entities* are things that have discrete, independent existence. An entity's existence does not depend on any other thing. *Associations* are things whose existence depends on one or more other things, such that if any of those things ceases to exist, then the thing itself ceases to exist or becomes meaningless. An associative database comprises two data structures: (i) A set of items, each of which has a unique identifier, a name and a type. (ii) A set of links, each of which has a unique identifier, together with the unique identifiers of three other things, that represent the source, verb and target of a fact that is recorded about the source in the database. Each of the three things identified by the source, verb and target may be either a link or an item.

The associative model is very similar to RDF semantics (described later in section 5.2.4). The idea of both approaches consists in representing semantics by triples subject-predicate-object, which are elementary data units. Subject is an element, which is characterized by object while the predicate designates the characteristic. The difference is that the associative model is used to represent semantics in databases while RDF is used to represent information on the web. One feature of this approach is that it essentially does not distinguish between different types of descriptive elements such as items (objects) and attributes (variables, properties). All elements are simply entities and their role is determined by their use in statements (associations). Another feature is that new statements (sentences) may involve both entities and associations.

Entity-Attribute-Value data model

The next approach, which tries to generalize and simplify data representation is the Entity-Attribute-Value (EAV) model of data representation [3]. The best way to understand the rationale of EAV design is to understand row modeling (of which EAV is a generalized form). Consider a supermarket database that must manage thousands of products and brands, many of which have a transitory existence. Here, it is intuitively obvious that product names should not be hard-coded as names of columns in tables. Instead, one stores product descriptions in a Products table: purchases/sales of individual items are recorded in other tables as separate rows with a product ID referencing this table. Conceptually an EAV design involves a single table with three columns, an entity (such as an olfactory receptor ID), an attribute (such as species, which is actually a pointer into the metadata table) and a value for the attribute (e.g., rat). In EAV design, one row stores a single fact. In a conventional table that has one column per attribute, by contrast, one row stores a set of facts. EAV design is appropriate when the number of parameters that potentially apply to an entity is vastly more than those that actually apply to an individual entity.

This approach is similar to associative model and RDF semantics in the sense that it breaks the whole representation into a set of primitive facts. Indeed the whole database is represented by only one table in EAV and by two tables in associative model.

Microsoft WinFS data model

The Microsoft WinFS [4] is a storage subsystem tightly integrated with the next version of Microsoft Windows operating system (Longhorn). It is an integrated approach, which combines features of many different methods such as relational, object-oriented and XML-based APIs.

The WinFS data model is designed on the concept of collections of items called *sets*. Items in WinFS are the primary data containers. Each item contains multiple properties to hold the data, and a reference to a type that defines what properties the item has. An item may contain additional properties that are not defined by its type.

A *type* is a schema that defines the structure of an item, relationship, or extension. Types may refer to other types to form an inheritance tree. Each item, relationship, or extension of a given type contains the properties defined by the type, as well as the properties defined by the type's parents. A given type can be used with items, relationships, or extensions, referred to as *item types*, *relationship types*, and *extension types*. WinFS contains a set of common schemas called *Windows types* — for example, Contacts, Sound Tracks, Messages, and Pictures. If the Windows types don't meet your application needs, with WinFS you can customize, or extend, a type. Applications can also create new types.

A *relationship* is an association between two items — the *source* and the *target*. Like an item, a relationship has a type that defines the required structure of its data, and may have additional properties attached.

*Properties* are the smallest data containers. A property consists of a PropertyName (used to refer to a particular property in a collection of properties) and a value. The value may be a primitive type (such as integer, string, or filestream), or it may be more complicated. A PropertyName is used to refer to a particular property in a collection of properties.

The main high level purpose of this new data model and storage is to let users categorize their data, and then filtering their view of the information by criteria that they assign. For example, a user could organize music by song title, artist, album, year released, track, song duration, or genre, and then filter the view of the matching music so that they see only songs titles beginning with the letter "L." With WinFS users can access anything they need by familiar contexts rather than by digging through a hierarchical folder tree.

## 5.2 Conceptual Models

### 5.2.1 Formal Concept Analysis

The goal of the Formal Concept Analysis (FCA) consists in developing methods for representation and analysis of data [5]. However, the specific feature is that the data is structured into units, which are formal abstractions of concepts of human thoughts allowing meaningful and comprehensive interpretation. In this sense FCA has the same goal as many other general approaches such as Ontologies, Semantic Web, Rough Sets and including the concept-oriented model described in this paper.

FCA is heavily based on the lattice theory and essentially it can be viewed as an adaptation of lattice theory to the needs of data representation and analysis. The primary elements of FCA are a set of *attributes M* and a set of *objects G*. Each object is characterized by a subset of attributes. Thus attributes can be considered Boolean variables so that each object takes either value 1 or 0. If an object takes value 1 of an attribute then it is said to be characterized by this attribute. In FCA this information is encoded in the form of *incidence relation* $I \in G \times M$. A triple $\langle G, M, I \rangle$ is called a *formal context*.

Here we see the first major difference between FCA and the concept-oriented model. In FCA the semantics is represented by the incidence relation, which explicitly encodes the relationship between objects and attributes they are characterized by. (In general case the incidence relation must not be Boolean.) In the concept-oriented approach we use references as the primitive representation mechanism, which allows us to implement attribute-value or item-subitem relation among items. Since this is a primitive mechanism there is no special table or other dedicated representation tools used to encode this relationship. We simply suppose that each item can be characterized by other items and this is the basic property of each individual item rather than a global relation among items and attributes. In particular, this allows items to play a role of attribute values for other items. In contrast in FCA attributes are special elements, which play special role. In other words, objects in FCA are characterized by attributes (represented by incidence relation) while in the concept-oriented model items are characterized by other items (represented by references).

Objects in FCA are analogous to items in the concept-oriented model. In FCA objects are defined as a primary element along with attributes and incidence relation. In CODBM items are defined as a part of semantics and each item is defined as a combination of some other items. The next big difference between these two approaches is that from the very beginning items are living in different concepts while FCA objects are living in one global set.

The third difference is that attributes in the concept-oriented model are not defined as special elements. Instead attributes are equivalent to concepts (sets of items). Thus we may have complex attributes or primitive attributes but all of them are concepts. In FCA the situation is more involved. Initially we have special elements called attributes, which might be thought of as primitive variables of Boolean type.

A pair $\langle A, B \rangle$ is a *formal concept* of $\langle G, M, I \rangle$ if and only if $A \in G$, $B \in M$ and they satisfy a certain property described below. $A$ is a subset of object and is called the *extent* of the concept $\langle A, B \rangle$. $B$ is a subset of attributes and is called the *intent* of the concept $\langle A, B \rangle$. Notice that the intent of a concept is defined as a set of attributes while in the concept-oriented model it is a set of other concepts. The extent is defined as subset of all objects while in CODBM the extent of a concept is not selected from some set but rather is defined by itself so it is an analogue of the whole set of objects $G$ in FCA.

In principle, any pair of some subset of objects and some subset of attributes might be a concept. A property that makes such an arbitrary pair $\langle A, B \rangle$ a concept is that $B$ is precisely the set of attributes characterizing objects from $A$, and dually, $A$ is precisely the set of objects from $G$ covered by the specified attributes from $B$. This property can be viewed as follows. If we add a new object to the concept extent then it entails the necessity to add one more attribute (otherwise it is not a concept). And dually, if we remove some attribute from the concept intent then it entails the necessity to remove also some objects (otherwise it is not a concept). In this sense concepts are optimal pairs, which can be used to represent the incidence relation. There are also other interpretations of concepts and their properties. In particular, there is a correlation in methods and notions with Boolean algebra.

It is very important that all formal concepts are naturally ordered by the subconcept-superconcept relation. Superconcepts cover more objects by means of fewer attributes. In particular, the top concept is the most general and covers all objects using no attributes at all. Concepts including a single attribute in its intent cover only objects, which are characterized by this attribute and so on. The ordered set of all formal concepts is called the *concept lattice*.

The fourth difference between these two approaches is that FCA defines formal concepts proceeding from the properties of the considered formal context. In other words, it is important that the number of and the structure of concepts depend on the syntax and semantics of the formal context. For example, if we add, delete or update an object or attribute in the formal context then the concepts may well change significantly. In contrast, concepts in CODBM are fixed in the database syntax or schema rather than are floating. This is because concepts in CODBM are considered generalization of attributes and since attributes are initially fixed (both in FCA and CODBM) all more complex attributes (concepts in CODBM) should be also fixed. Generally it is quite natural that normally we tend to define the syntax or structure of any problem domain and then fill it with some semantics.

Another general difference is that in CODBM we have an explicit separation between syntax and semantics and particularly there exists two lattices: that of concepts and that of items. The lattice of concepts define the model syntax, i.e., a hierarchical multidimensional system of coordinates where objects may live. The lattice of items represents the model semantics, i.e., a set of really existing objects selected among all possible objects. In FCA there is not such an explicit separation and the model is represented by the formal context, which consists of attributes, objects and the incidence relation. The attributes could be viewed as a syntax but in contrast to CODBM it is not hierarchical so the attributes in FCA correspond to primitive concepts in CODBM. The objects and the incidence relation could be viewed as the semantics. The semantics is broken into two parts because it makes the theoretical analysis significantly easier. Particularly, many formalisms (lattice theory, ordered sets, Boolean algebra etc.) formulate their results in the form, which is close to incidence relation between objects and attributes.

The FCA representation can be obtained from CODBM by transforming it to canonical form and removing item references. In this case there is only one bottom concept items of which are characterized by primitive concepts. The bottom concept is considered as a set of objects $G$, primary concepts are multi-valued attributes. We can further transform multi-valued attributed to binary attributes. For that purpose each primitive item has to be considered a primitive concept with two values. Then each item from the bottom concept will be characterized by some subset of all primitive binary concepts.

### 5.2.2 Entity-Relationship Model (ERM)

E-R diagrams can be used to express data requirements in business terms and can be used to document and analyse business rules [6]. The same kind of diagrams can be used by IT professionals and systems analysts to understand and document systems data requirements technical implementation designs.

One of the main problems of the relational model is that it does not have relationships. Indeed we can create tables and then manipulates records within these tables but there is no mechanism to represent higher level relationships between records. The entity-relationship model tries to fill this gap by considering explicitly two main elements: entities and relationships. Entities are supposed to represent objects of the real world while relationships are elements designating existing relations among these objects. It is very convenient to represent relationships as arrows connecting entities. Such an approach could overcome the problem of relational model by allowing direct modelling of relationships. However, since relationships cannot be represented directly in the database but rather are transformed into some table records this approach is viewed as a conceptual one. In other words, any entity-relationship model has to be transformed into the corresponding relational model where relationships are records or record properties.

The entity-relationship model tries to solve the fundamental problem of representation of relationships between objects. We know how we can represent objects but it is difficult to represent relations among them because relations are known to be fundamentally different from objects. The concept-oriented model proposes a general solution to this problem by assuming that relations between objects are also objects having lower level. So in the concept-oriented model we have a lattice of concept as the underlying structure where items from subconcepts can be treated as instances of relations between the superconcepts. In the concept-oriented approach there is only one basic relation — inclusion relation implemented via upward references. Instances of this relation are not items of any concept in the database schema (they are supposed to exist as object on physical levels of database implementation). Although other approaches also translate relations into objects, the difference is that the concept-oriented model uses it as a primary principle.

### 5.2.3 Multidimensional Databases and OLAP

OLAP and multidimensional databases [7] are not normally considered an independent database model. Rather these technologies represent significant conceptual shift in understanding what is a database and how it should be used, which has been developed during the last decade. Obviously, this change of paradigm is not spontaneous and is based on the real needs of both industry and research. This shift in paradigm is in fact very close to what is considered in the concept-oriented model. Indeed, two main things considered in OLAP — multidimensionality and hierarchy — are precisely what is modelled formally in CODBM. Below we describe some differences between these two approaches:

- In OLAP there is no formal semantics, i.e., it is only view on existing data and the way this data is obtained by means of aggregation.

- It can be viewed as a technique for changing view hierarchically in multidimensional space by means of drill down and roll up operations.

- In OLAP there is not a notion of inference or deduction. Instead of them a notion of aggregation is used (which is one of technical premises for formally defining inference in CODBM).

- The structure is normally fixed in the sense that we have a number of dimensions and selected measures. It is difficult to change something, particularly, use measure as a dimension, because the structure of dimensions and measures is based on the corresponding predefined queries used to calculate the operation. In CODBM once the semantics has been defined we can then easier formulate complex queries and the database schema will guarantee consistency of the result.

### 5.2.4 RDF and Semantic Web

The Resource Description Framework (RDF) is a language for representing information about resources in the World Wide Web [8,9]. However, by generalizing the concept of a "Web resource", RDF can also be used to represent information about things that can be *identified* on the Web, even when they cannot be directly *retrieved* on the Web.

RDF is based on the idea that the things being described have properties, which have values (objects). In RDF each fact is represented by a triple Subject-Predicate-Object. The subject is the part that identifies the thing the statement is about. The predicate is the part that identifies the property or characteristic of the subject that the statement specifies. The object is the part that identifies the value of that property.

Each triple represents a statement of a relationship between the things denoted by the nodes that it links. The assertion of an RDF triple says that some relationship, indicated by the predicate, holds between the things denoted by subject and object of the triple. A familiar representation of such a fact might be as a row in a table in a relational database. The table has two columns, corresponding to the subject and the object of the RDF triple. The name of the table corresponds to the predicate of the RDF triple. A further familiar representation may be as a two place predicate in first order logic.

The main difference between RDF and CODBM is their interpretation of relationships and properties. (The way in which relationships and properties are treated is one of the main distinguishing features of any representation mechanism.) In RDF it is supposed that relationships are equivalent to variables (dimensions, properties), i.e., this mechanism does not distinguish between properties and relationships, which are supposed to be equivalent. This is why if we identify a relationship between entities it is encoded as a triple with the entity name, relationship name and the value name.

In contrast, In CODBM we follow the fundamentally different approach where relationships among items are distinguished from characterization mechanism. The mechanism of item characterization has a primary role. This mechanism is implemented via item references and is interpreted as item inclusion, item characterization or some other basic relationship among items. The multidimensional hierarchy established via this characterization or inclusion relation determines the database semantics and is the basis for all other higher level relationships. In other words, all other relationships are derived from this basic relation between items. Thus we explicitly select the characterization (inclusion, attribute-value etc.) relation among items as a fundamental element and particularly distinguish it from all other custom relationships, which can be derived from it. The main property of this basic relation is that we do not have its instances available (at this level). In other words, the inclusion relation instances are not items with their own references and properties (characterization).

The higher level relations among items have a secondary role and do not directly define the data semantics. They have custom nature and are defined for convenience purposes. In other words, we can define the data semantics using only the basic inclusion relation or we can define higher level custom relationships and then use them to define the data semantics. If the semantics is already defined via the basic inclusion relation then we can derive custom higher level relationships using some definition. Or we can use such definition of relationships to change the data semantics by adding, removing or updating the relationship instances. In this case however the relationship instances are real objects with their references. Thus any definition of higher level relationship among items is simply translated into primitive operations on items and their characteristics. Notice again that in RDF there is only one mechanism of subject characterization with no separation between inclusion relation and higher level custom relationships. In CODBM we proceed from the principle "relation as an item at lower level", which means that any relation is translated into a set of normal items, which in turn can use other items lower in the hierarchy to define its own relations. Such a translation proceeds down the lowest level of representation where we use inclusion relation the instances of which are not items (they are still items but existing at physical level of representation, i.e., they are not accessible from the database).

Another distinction between RDF and CODBM is their formalization of set-valued properties. RDF uses for that purpose a dedicated mechanism (like most other approaches). This mechanism is designed to describe a group of things. For example, it might be needed to say that a book was created by several authors, or to list the students in a course, or the software modules in a package. RDF provides several predefined (built-in) types and properties that can be used to describe such groups. In CODM we implement set-valued properties by means of the basic mechanism, i.e., set-valued properties is a convenient abstraction or an additional layer over the basic representation mechanism. Here CODM proceeds from the principle that there is no separate notion of multi-valued property but it is simply one view on the semantics defined via single-valued attribute-value relation (interpreted also as an inclusion or characterization relation and implemented via references).

### 5.2.5 Ontologies

Ontology and semantic web is first of all an aggregated effort, which combines a lot of different approaches and methods aimed at bringing conceptual semantics and structure into different areas. Particularly, all these approaches are highly rich and expressive and look more like high level representation and description languages. Yet the main idea is that an initially unorganised set of objects can be structured by using different relations and mechanisms like properties and hierarchies.

An ontology is an explicit formal specification of the terms in the domain and relations among them [10]. An ontology defines a common vocabulary in order to share information in a domain. It includes machine interpretable definitions of basic concepts in the domain and relations among them. Classes are the focus of most ontologies. Like in many other approaches (including the concept-oriented model) classes describe concepts in the domain and many have subclasses, which describe more specific concepts. Slots describe properties of classes and instances. Developing an ontology normally includes the following steps: (i) defining a set of classes in the ontology, (ii) establishing a subclass-superclass hierarchy among them, (iii) defining slots for classes, (iv) defining instances of classes by filling in their slot values.

Below we describe main differences between ontology engineering and CODBM:

- Ontology is in great extent an informal paradigm. Particularly, there is normally no one correct way to model a domain and developing ontology is in most cases an iterative process. CODBM is a formal representation mechanism, which concretely and constructively defines its elements and procedures.

- Ontology uses inheritance (extension, specialization, subclass-superclass) relation as the main type of relationships among concepts. In this sense it is similar to object-oriented and some other paradigms. In CODBM we do not use inheritance. Instead we define each new subconcept as a combination of a number of superconcepts and then each item is equal to a combination of its superitems.

- Ontology is mainly intended to represent terms in the domain. CODBM explicitly separates the syntax and semantics where syntax consists of concepts and their dimensions while semantics consists of items defined via their superitems. Higher level superitems have longer life-cycle and represent general terms and once they have been defined can serve as characteristics of new lower level subitems, which are more specific and have shorter life-cycle. So these items can be considered common terms or shared knowledge. Concept structure is used to organize this knowledge. It is important to notice that in contrast to ontologies, concepts in CODBM are not terms used to describe the reality. Instead CODBM uses higher level items to characterize and describe lower level items. Concepts in this case are used as organizational units providing a structure for the space where items live. In other words, we might put all items in one large space (concept) where they will have high dimensionality but normally want to introduce more intermediate concepts each with lower dimensionality. The concepts in CODBM play a role of hierarchically organized spaces and this role is definitely different from the role of concepts in ontology engineering where they are used as terms themselves to describe the reality.

- Ontologies use characterization via slots while in CODBM all items are characterized via other items (from superconcepts). Elementary characteristics or properties are primitive concepts. Thus concept hierarchy (syntax or schema) development in CODBM can be thought of as defining new complex attributes via already existing attributes because each concept can be used to characterize other concepts.

- Ontologies are heavily based on the object-oriented paradigm and so this approach has the same specific features and differences with the concept-oriented model. In particular, ontologies have the same inheritance mechanism where base objects loose their identity.

# 6 Conclusions

In this paper we described the concept-oriented data model. Below we summarize main properties of this model:

- Explicit separation between syntax and semantics, which have their own separate structure.

- Hierarchical multidimensional structure for concepts, which is interpreted as a hierarchical attribute system.

- Formal representation of the data semantics and its transformations including the canonical form.

- Data items are defined via other data items as their combination.

- Interpretation of the semantics as different types of relationships and mechanisms such as hierarchical inclusion of items into sets, hierarchical attribute value characterization, item coordinates and hierarchical positioning.

- Interpreting items as relation instances for their parent items.

- Strict reference structure where items are allowed only to store references to their parent items. References are the mechanism of implementation of data semantics rather than an arbitrary representation and access method.

- Multi-valued attributes are a view on the basic semantics rather than a built-in mechanism.

- Life-cycle management via null propagation and use control (reference counting and garbage collection). Special role for null value as the absence of parent for an item.

- Semantics propagation through the schema rules and their use for such mechanisms as aggregation, virtual properties, OLAP, inference.

- Query language based on collection manipulation.